\documentclass[aps,notitlepage,onecolumn,nofootinbib,superscriptaddress,longbibliography,11pt]{revtex4-2}

\usepackage[nodisplayskipstretch]{setspace}

\everydisplay=\expandafter{\the\everydisplay\setstretch{1}}

\usepackage{graphicx}
\usepackage{dcolumn}
\usepackage{bm}
\usepackage[table]{xcolor}
\usepackage{booktabs}
\usepackage[utf8]{inputenc}
\usepackage[T1]{fontenc}
\usepackage{lineno}

\usepackage[normalem]{ulem}
\usepackage{amsthm}
\usepackage{amssymb}
\usepackage{mathtools} 
\usepackage{bm}
\usepackage{braket}
\usepackage{lineno}
\usepackage{pifont}
\usepackage{bbm}
\usepackage{bbding}
\usepackage{hyperref}
\usepackage{algorithm} 
\usepackage{algpseudocode} 
\usepackage{graphicx}
\usepackage{dcolumn}
\usepackage{bm}
\usepackage{subfigure}
\usepackage{comment}
\usepackage{tikz}
\usepackage{amsfonts}
\usepackage{physics}
\usepackage{amsmath}
\usepackage{amssymb}
\usepackage{mathrsfs}
\usepackage[capitalize]{cleveref}
\usepackage{accents}
\usepackage{lineno}
\allowdisplaybreaks

\newtheorem{theorem}{Theorem}
\newtheorem{definition}{Definition}
\newtheorem{proposition}{Proposition}
\newtheorem{corollary}[proposition]{Corollary}
\newtheorem{lemma}{Lemma}

\newcommand{\boldtheta}{\boldsymbol{\theta}}

\newcommand{\boldf}{f}

\newcommand{\bfx}{\mathbf{x}}
\newcommand{\KL}{\mathrm{KL}}

\newcommand{\ubar}[1]{\underaccent{\bar}{#1}}

\newcommand{\cP}{{\mathcal{P}}}
\newcommand{\cPN}{{\cP^{\otimes N}}}
\newcommand{\cQ}{{\mathcal{Q}}}
\newcommand{\cO}{{\mathcal{O}}}
\newcommand{\cA}{{\mathcal{A}}}
\newcommand{\cE}{{\mathcal{E}}}
\newcommand{\cX}{{\mathcal{X}}}
\newcommand{\cK}{{\mathcal{K}}}
\newcommand{\cH}{{\mathcal{H}}}
\newcommand{\cY}{{\mathcal{Y}}}
\newcommand{\bP}{{\mathbb{P}}}

\newcommand{\bE}{{\mathbb{E}}}
\newcommand{\bI}{{\mathbb{I}}}
\newcommand{\RY}{\mathrm{R}_\mathrm{Y}}
\newcommand{\RZ}{\mathrm{R}_\textbf{}{Z}}

\newcommand{\Rz}{\mathrm{R}_{\mathrm{Z}}}
\newcommand{\cRz}{\mathrm{C-{R}}_{\mathrm{Z}}}

\begin{document}
	
	\preprint{APS/123-QED}
	
	\title{Near-optimal Prediction Error Estimation for Quantum Machine Learning Models}
	
	\author{Qiuhao Chen}
	\affiliation{Yangtze Delta Industrial Innovation Center of Quantum Science and Technology, Suzhou 215000, China}
	\affiliation{
		School of Mathematics and Statistics, Wuhan University, Wuhan 430072, China 
	}
	\author{Yuling Jiao}
	\affiliation{
		School of Artificial Intelligence, Wuhan University, Wuhan 430072, China 
	}
	\affiliation{National Center for Applied Mathematics in Hubei, Wuhan 430072, China}
	\affiliation{Hubei Key Laboratory of Computational Science, Wuhan 430072, China}
	\author{Yinan Li}
	\affiliation{
		School of Artificial Intelligence, Wuhan University, Wuhan 430072, China 
	}
	\affiliation{National Center for Applied Mathematics in Hubei, Wuhan 430072, China}
	\affiliation{Hubei Key Laboratory of Computational Science, Wuhan 430072, China}
    \affiliation{Wuhan Institute of Quantum Technology, Wuhan, 430072, China.}
	\author{Xiliang Lu}
	\affiliation{
		School of Mathematics and Statistics, Wuhan University, Wuhan 430072, China 
	} 
	\affiliation{National Center for Applied Mathematics in Hubei, Wuhan 430072, China}
	\affiliation{Hubei Key Laboratory of Computational Science, Wuhan 430072, China} 
	\author{Jerry Zhijian Yang}
	\affiliation{Institute for Math \& AI, Wuhan, Wuhan 430072, China}
	\affiliation{
		School of Mathematics and Statistics, Wuhan University, Wuhan 430072, China 
	}
        \affiliation{National Center for Applied Mathematics in Hubei, Wuhan 430072, China}
	\affiliation{Hubei Key Laboratory of Computational Science, Wuhan 430072, China}
	\date{\today}
	
	\begin{abstract}
		Understanding the theoretical capabilities and limitations of quantum machine learning (QML) models to solve machine learning tasks is crucial to advancing both quantum software and hardware developments. Similarly to the classical setting, the performance of QML models can be significantly affected by the limited access to the underlying data set. Previous studies have focused on proving generalization error bounds for any QML models trained on a limited finite training set. We focus on the optimal QML models obtained by training them on a finite training set and establish a \emph{tight} prediction error bound in terms of the number of trainable gates and the size of training sets. To achieve this, we derive covering number upper bounds and packing number lower bounds for the data re-uploading QML models and linear QML models, respectively, which may be of independent interest. We support our theoretical findings by numerically simulating the QML strategies for function approximation and quantum phase recognition.
	\end{abstract}
	
	
	\maketitle
	
	\section{Introduction}
	Quantum machine learning (QML) is an emerging field that combines machine learning principles with recent advancements in quantum technology~\cite{biamonte2017quantum, Schuld15IntroQML,cerezo2021variational,ciliberto2018QuantumMachineLearning, schuld2021MachineLearningQuantum,RevModPhys.94.015004,Schuld2019QMLinFeatureHilbertSpace,zhu2019TrainingQuantumCircuits,PerezSalinas2020datareuploading,Dunjko_2018,Tian2023QNNGeneLearn,li2021RecentAdvancesQuantum,cerezo2022ChallengesOpportunitiesQuantum}. The main idea of QML is to train a QML model consisting of parameterized quantum circuits using classical data, in order to accomplish machine learning tasks. 
	QML has exhibited enhanced expressive power and has great potential to achieve quantum advantages on specific learning tasks~\cite{abbas2021PowerQuantumNeural,du2023ProblemDependentPowerofQNN,wang2021towards,Jerbi2021QuantumEnhancementsDRL,Pirnay2023SuperpolynomialQuantumClassicalSeparation}
	In particular, recently proposed QML strategies can adapt to the limitations of near-term quantum devices~\cite{havlivcek2019supervised,mcclean2016theory, endo2021hybrid, farhi2014quantum, doi:10.1126/science.abn7293, huang2022provably,peruzzo2014VariationalEigenvalueSolver,moll2018quantum,coyle2020BornSupremacyQuantum,Rudolph2022GenerationHighResolutionHandwritten,pan2023DeepQuantumNeural,pan2023ExperimentalQuantumEndtoend,ren2022ExperimentalQuantumAdversarial,zhang2022ExperimentalDemonstrationAdversarial}. Thus, understanding the capabilities of QML is essential for both quantum software and hardware developments. 

	The goal of quantum-supervised learning is to train a QML model using classical labeled data to predict the labels of unseen data. Mathematically, we are looking for an optimal QML model on a training data set to approximate the real data-label mapping satisfying certain functional properties. In this situation, the universal approximation theorem for QML models plays an important role: It claims that QML models with sufficiently large size should be sufficient to approximate general function classes, such as Lipschitz continuous functions and smooth functions~\cite{schuld2021effect,vidal2004KAK,perez-salinas2021one,NEURIPS2022_b250de41,manzano2023parametrized,goto2021universal,gonon2023universal,yu2024,qiTheoreticalErrorPerformance2023}.  
	Considering the training process, the performance of QML models can be affected by many practical issues, such as the quality of the quantum hardware, the efficiency of training algorithms, and the size of the training data sets. 
	\begin{figure}[ht!]
		\center
		\includegraphics[width=.5\textwidth]{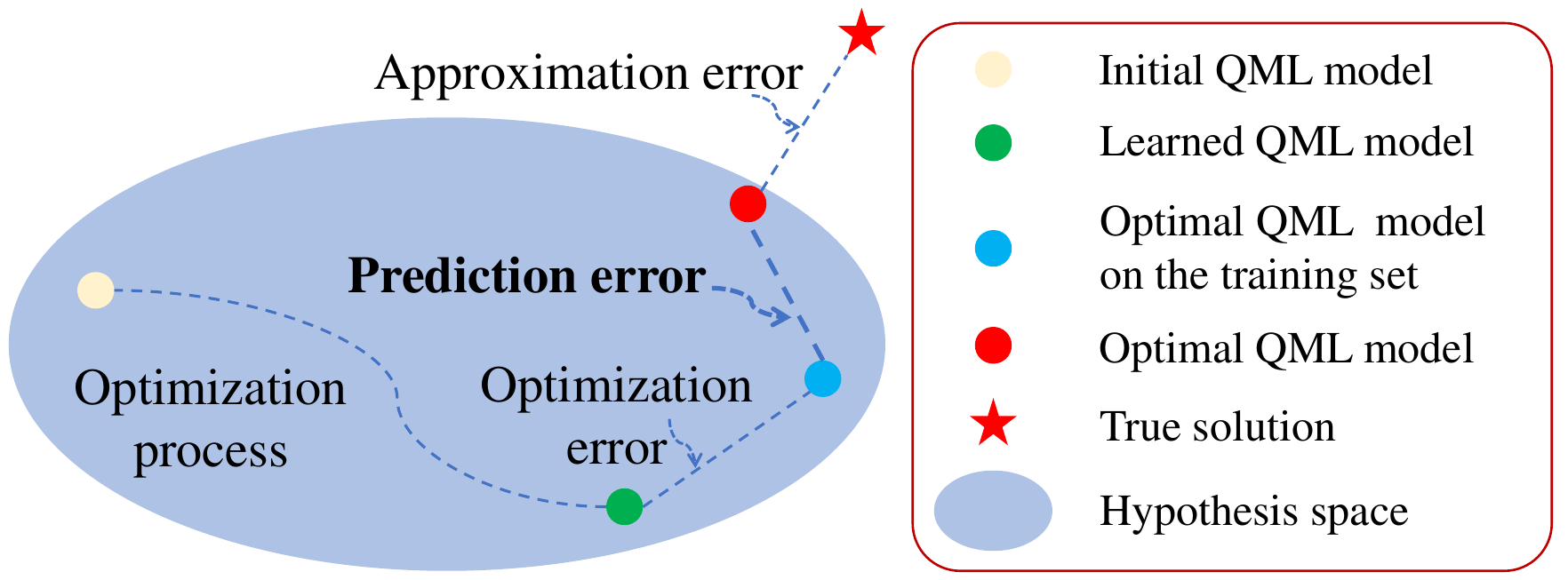}	
		\caption{\small{Performance analysis of QML models. The QML process begins with an initial model (white bullet) and progresses to a learned QML model (green bullet) through a hybrid optimization procedure, such as variational quantum algorithms. The distance between the learned QML model and the optimal QML model on the training set (blue bullet) is referred to as the {optimization error}, which quantifies the performance of the optimization procedure. The prediction error quantifies the gap between the optimal QML model on the training set and the optimal QML model (red bullet), measuring the influence resulting from a limited training set. The approximation error, representing the distance between the optimal QML model and the target function (star bullet), reflects the ability of the hypothesis space generated by the QML model to approximate the target functions.}}
		\label{fig: Q learning process}
	\end{figure}
	As mentioned in~\cite{qiTheoreticalErrorPerformance2023,yu2024} and illustrated in FIG.~\ref{fig: Q learning process}, the theoretical performance analysis of QML models can be understood through (1) the \emph{optimization error}, which quantifies the quality of the QML training process; (2) the \emph{prediction error}, which assesses the capability of predicting the unseen data; and (3) the \emph{approximation error}, which measures the gap between the optimal QML models with the real data-label pair mappings. Comprehending each error term is essential in understanding the theoretical performance of QML models. 
	
	Note that the optimization error depends greatly on the training strategies, especially the classical optimization algorithms, and the approximation error can be estimated through quantitative universal approximation theorems of QML models. The prediction error should be understood from both computational and physical perspectives. It is clear that the more training data we use, the more accurate the prediction will be. On the other hand, if we utilize a big training data set, uploading the training data into the QML models could be a challenging task; if a small training data set is used, then we can merely hope to use the optimal QML model on such a small training set for prediction. Thus, understanding the interplay between the prediction error of QML models, the training data set size (\emph{sample complexity}), and the structural properties of QML models (\emph{model complexity}) is of great importance. 

	In this work, we focus on bounds on the prediction error of QML models in terms of the sample complexity and the model complexity of QML models. Previous works on estimating the prediction error of QML models have focused on understanding their generalization error~\cite{caroGeneralizationQuantumMachine2022,qiTheoreticalErrorPerformance2023,9517721,jerbiQuantumMachineLearning2023a,Bu_2023,bu2021RademacherComplexityNoisy,bu2022StatisticalComplexityQuantum,du2022EfficientMeasureExpressivity,gyurik2023StructuralRiskMinimization,schatzki2024TheoreticalGuaranteesPermutationequivariant,cai2022SampleComplexityLearning,chung_et_al2021SampleEfficient}. Roughly speaking, generalization error measures the difference between the average loss of any QML models on a given training data set and the average loss on the whole data set. \citet{caroGeneralizationQuantumMachine2022} developed a comprehensive framework for analyzing the generalization error and proved that it scales linearly with $\sqrt{T\log T/N}$. Following this, significant efforts have been dedicated to extending the generalization bound of QML models from different perspectives, such as the noise level and the underlying distributions of data sets~\cite{caro2020PseudodimensionQuantumCircuits,PRXQuantum.2.040321,caroOutofdistributionGeneralizationLearning2023,Peters2023generalization,caroinformation,caro2021encoding,huangcorset,haug2023GeneralizationQuantumGeometry,fuster2023UnderstandingQuantumMachine,qian2022TheDilemma}.
	
	We mention that the generalization error bound can be adopted to upper bound the prediction error (see Lemma~\ref{lem: gen to ER}). However, the generalization-error-type bounds are too general to bound the prediction error. Recent numerical experiments indicate that the general bounds cannot explain QML models' behavior in specific tasks, such as solving quantum phase recognition~\cite{fuster2023UnderstandingQuantumMachine}. This is similar to the classical setting, where generalization error bounds failed to explain the success of deep convolutional neural networks~\cite{10.1145/3446776}. 

    We prove near-optimal bounds on the prediction error of QML models in terms of the sample complexity and the model complexity of QML models. 
    In particular, if we utilize a training data set of size $N$ to train a QML model with $T$ trainable quantum gates, then \emph{the prediction error scales linearly with $T/N$}--a quadratic improvement over previous results based on generalization error. We also prove a matching lower bound for the prediction error by exhibiting supervised learning tasks such that the prediction error of any QML model scales at least linearly with $T/N$. Such a bound is also meaningful from a practical perspective: If we have a prior estimate of the required size of the training set and the structure of QML models, we can optimize the data-encoding quantum circuits in advance to meet the limitations of near-term devices. This would help implement QML strategies on near-term devices with theoretical performance guarantees.

	To establish our prediction error bound, we utilize advanced tools from statistical theory and Bayesian analysis, which have also played an important role in classical deep learning theory. In particular, we prove matching packing lower bound and covering upper bound for QML models, which might be of independent interest. We numerically simulate the performances of two fundamental QML tasks, using single-qubit parameterized quantum circuits to approximate univariate analytic functions~\cite{NEURIPS2022_b250de41} and employing quantum convolutional neural networks to recognize symmetry-protected topological phase~\cite{congQuantumConvolutionalNeural2019}. The relations among their prediction error, the training set size, and the number of trainable gates validate our theoretical findings.

	\section{Results}
	\paragraph*{Problem formulation.} The QML models considered throughout are of \emph{data re-uploading} type~\cite{PerezSalinas2020datareuploading}, i.e., consisting of interleaved data encoding circuit blocks and trainable circuit blocks.
	More precisely, let $\bfx\in[0,1]^n$ be the input data vector and $\boldtheta = (\bm{\theta}_1, \ldots, \bm{\theta}_L)\in (0,2\pi]^T$ be a set of trainable parameter vectors, where $T$ is the total number of trainable parameters, and each $\bm{\theta}_i$ represents the trainable parameters in the $i$th layer. Let $U_i(\boldtheta_i)$ be an $n$-qubit trainable quantum circuit with trainable parameter vector $\bm\theta_i$ and let $V(\bfx)$ be prefixed data encoding quantum circuit for encoding the classical data $\bfx$. An $L$-layer data re-uploading PQC can be then expressed as
	\[
	U_{\bm\theta}(\bfx) = U_L(\boldtheta_L) V(\bfx) U_{L-1}(\boldtheta_{L-1})\cdots U_1(\boldtheta_1)V(\bfx).
	\]
	Applying $U_{\bm\theta}(\bfx)$ to a quantum state and measuring the output states provides a way to express functions on $\bfx$:
	\[
	\boldf_{\bm\theta}(\bfx) \coloneqq \bra{\mathbf{0}^n} U^\dagger_{\bm\theta}(\bfx) \cO U_{\bm\theta}(\bfx) \ket{\mathbf{0}^n}, 
	\]
	where $\cO$ is some Hermitian observable. When $L=1$, we call such models \emph{linear QML models}~\cite{jerbiQuantumMachineLearning2023a}. Let $\cH=\{\boldf_{\boldtheta}(\bfx):~{\boldtheta}\in\mathbf{\Theta}\}$ be the hypothesis space of the data re-uploading QML model. 
	
	Supervised learning tasks ask to find a suitable parameter $\boldtheta^*$ such that $\boldf_{\boldtheta^*}(\bfx)\approx y$ for any $(\bfx,y)\in\mathcal X\times \mathcal Y$, where $\bfx\in\cX\subseteq\mathbb{R}^{n}$ is the sample vector and $y\in\cY\subseteq\mathbb{R}$ is the label of $\bfx$. For any $\boldtheta\in{\mathbf{\Theta}}$ and $(\bfx,y)\in\mathcal X\times\mathcal Y$, we utilize the \emph{$\ell^2$ loss function} $\ell(\boldf_{\boldtheta}; \bfx, y):= (\boldf_{\boldtheta}(\bfx)-y)^2$ to assess the performance of $\boldf_{\boldtheta}$ at $(\bfx,y)$. Let $\cP$ be the probability density function of $\mathcal{X}\times\mathcal{Y}$. The average loss (also known as the population risk) of a hypothesis function $\boldf_{\boldtheta}$ is defined as
    \begin{eqnarray}\label{eqn:expected_loss}
    R(\boldf_{\boldtheta}):=\mathbb{E}_{(\bfx, y)\sim \cP}\left[ \ell(\boldf_{\boldtheta}; \bfx, y) \right],
    \end{eqnarray}
    The \emph{optimal QML model} $\boldf_{\boldtheta^*}$ is computed by minimizing the average loss, i.e.~${\boldtheta^*}\in\mathop{\arg\min}_{\boldtheta\in \mathbf{\Theta}} R(\boldf_{\boldtheta})$. From an optimization viewpoint, the performance of any hypothesis function $\boldf_{\boldtheta}$ can be evaluated by the so-called \emph{prediction error}, defined as the difference between the average loss of $\boldf_{\boldtheta}$ and $\boldf_{\boldtheta^*}$:
  	\begin{equation}\label{eq: prediction error}
    \cE_P(\boldf_{\boldtheta}):=R(\boldf_{\boldtheta})-R(\boldf_{\boldtheta^*}).
  	\end{equation}
	
	Since the probability density $\cP$ is usually unknown, the prediction error is hard to compute. The conventional approach is to utilize a finite training set $S=\{(\bfx_1, y_1),\dots,(\bfx_N, y_N)\}\subseteq\mathcal X\times\mathcal Y$ of size $N$, where each training data is sampled from $\cP$ independently. The average loss of a hypothesis function $\boldf_{\boldtheta}$ on the training set $S$ (as known as the empirical risk) is defined as 
	\begin{eqnarray}\label{eqn:empirical_risk}
    \hat{R}_S(\boldf_{\boldtheta}) := \frac{1}{N}\sum_{i=1}^{N}\ell(\boldf_{\boldtheta}; \bfx_i, y_i).
  	\end{eqnarray}
  	The \emph{optimal QML model $\boldf_{\hat{\boldtheta}_S}$ on the training set $S$} is computed by minimizing the average empirical loss, i.e.~$\hat{\boldtheta}_S\in\mathop{\arg\min}_{\boldtheta\in \mathbf{\Theta}} \hat{R}_S(\boldf_{\boldtheta})$. 

  	There are two critical questions relating to the above optimization problems: (1) Whether the average loss on the training set approximates the true average loss; and (2) Whether the optimal QML model on the training set approximates the optimal QML model. 
  	The first problem leads to the study of the generalization error of hypothesis functions, defined as
  	\begin{equation}\label{eq: generalization error}
    \text{gen}(\boldf_{\boldtheta}, S):=|R(\boldf_{\boldtheta})-\hat{R}_S(\boldf_{\boldtheta})|.
  	\end{equation}
  	\citet{caroGeneralizationQuantumMachine2022} proved that for any hypothesis function of linear QML models ($L=1$) with $T$ trainable quantum gates and a training set ${S}$ of size $N$ sampled according to the distribution $\cP^{\otimes N}$, the expected generalization error of $\boldf_{\boldtheta}$ is at most $O(\sqrt{T\log T/{N})}$. 

  	The generalization error bound can be utilized to answer (2) through the following lemma: 
  	\begin{lemma}\label{lem: gen to ER}
  	The expected prediction error (defined in~(\ref{eq: prediction error})) of the optimal QML models on a given training set (defined as the minimizer of~(\ref{eqn:empirical_risk}))is at most $O(\sqrt{T\log T/{N})}$. 
  	\end{lemma}
	\begin{proof}
	Let $\hat{\boldtheta}_S\in\mathop{\arg\min}_{\boldtheta\in \mathbf{\Theta}} \hat{R}_S(\boldf_{\boldtheta})$ be the empirical average loss minimizer. We have  
	\begin{equation*}
		\begin{split}
			&\bE_{S\sim \cP^{\otimes N}}[\cE_P(\boldf_{\hat\boldtheta_S})]
   \\=&\bE_{S\sim \cP^{\otimes N}}[R(\boldf_{\hat\boldtheta_S})-\hat{R}_S(\boldf_{\hat\boldtheta_S})+\hat{R}_S(\boldf_{\hat\boldtheta_S})-R(\boldf_{\boldtheta^*})]\\
		\leq& \bE_{S\sim \cP^{\otimes N}}[R(\boldf_{\hat\boldtheta_S})-\hat{R}_S(\boldf_{\hat\boldtheta_S})+ \hat{R}_S(\boldf_{\boldtheta^*})-R(\boldf_{\boldtheta^*})]\\
			=&\bE_{S\sim \cP^{\otimes N}}[\text{gen}(\boldf_{\hat\boldtheta_S})],
		\end{split}
	\end{equation*}
	\sloppy
	where the inequality holds since $\hat{R}_S(\boldf_{\hat\boldtheta_S})\leq \hat{R}_S(\boldf_{\boldtheta^*})$ and the second equality holds since \(\bE_{S\sim \cP^{\otimes N}}[\hat{R}_S(\boldf_{\boldtheta^*})]=R(\boldf_{\boldtheta^*})\).
	\end{proof}

	\paragraph*{Main results.} Note that the generalization error holds for any hypothesis function of QML models. Thus, the prediction error bounds in Lemma~\ref{lem: gen to ER} should not be tight. This was observed in~\cite{fuster2023UnderstandingQuantumMachine} by analyzing the prediction performance of quantum convolutional neural networks. Our main result is an \emph{improved prediction error bound} for the optimal QML models on a given training set:
	\begin{theorem}\label{thm:main_upper}
        For data re-uploading QML models with (at most) $T$ trainable quantum gates, the prediction error of the optimal QML models on the training set $S$ satisfies
		\begin{eqnarray}
			\bE_{S\sim\cP^{\otimes N}}\left[\cE_P(\boldf_{\hat{\boldtheta}_S})\right] = \tilde{O}\left(\frac{T}{N}\right).
		\end{eqnarray}
        where $\tilde{O}(\cdot)$ hides polylogarithmic factors. 

	\end{theorem}
	Theorem~\ref{thm:main_upper} \emph{quadratically improves} the prediction error upper bound obtained from generalization error bounds (cf.~Lemma~\ref{lem: gen to ER}), when the number of trainable parameters is fixed. In particular, the optimal QML model on a training data set of size $\Omega(T/\delta)$, instead of $\Omega(T\log T/\delta^2)$, is enough to achieve a $\delta$-prediction error.

	Our next result establishes a matching prediction error lower bound for \emph{linear QML models}. More precisely, we introduce a family of supervised learning problems called the \emph{Gaussian denoise problems} such that the expected prediction error of the optimal QML model on a sampled training data set of size $N$ cannot be smaller than $\Omega(T/N)$. 

	A Gaussian denoise problem asks to recover a parameter configuration $\boldtheta^*$ using noisy samples of the form $(\bfx,y_{\boldtheta^*})$, which satisfies
	\begin{equation}\label{eq: gaussian noise}
		y_{\boldtheta^*}=\boldf_{\boldtheta^*}(\bfx)+\varepsilon,
	\end{equation}
    where $\boldf_{\boldtheta^*}$ is an hypothesis function obtained from the linear QML model and $\varepsilon$ is a random Gaussian noise satisfying (1) $\mathbb{E}[\varepsilon]=0$, $\mathbb{E}[\varepsilon^2]=\sigma^2<\infty$ and (2) the noise is independent with $\bfx$. 
    Gaussian denoise problems have been extensively studied in deep learning theory and nonparametric regression (see e.g.~\cite{10.1214/23-AOS2266} and reference therein). It can be shown that $\boldf_{\boldtheta^*}$ is exactly obtained from the optimal linear QML model. Notable instances of Gaussian denoise problems in quantum computation include function approximation~\cite{schuld2021effect,NEURIPS2022_b250de41}, quantum compiling~\cite{Khatri2019quantumassisted}, quantum phase recognition~\cite{congQuantumConvolutionalNeural2019,Wu2023quantumphase}, and learning quantum dynamics~\cite{caroOutofdistributionGeneralizationLearning2023,PhysRevLett.126.190505,doi:10.1126/science.abn7293}. 
	
	We prove a minimax lower bound for the prediction error of linear QML models for solving the Gaussian denoise problems.
	\begin{theorem}\label{thm:main_lower}
	Let $\mathbb{A}$ be the set of all statistical QML strategies utilizing linear QML models with $T$ trainable quantum gates. 
 	Let $\cP_{\boldtheta}$ be the density of sample data pair $(\bfx, y_{\boldtheta})$, generated from the corresponding Gaussian denoise problem with unknown parameter $\boldtheta$. Then
	\[
	\inf_{\cA\in\mathbb{A}} \sup_{\boldtheta\in\mathbf{\Theta}} \mathbb{E}_{S\sim\cP_{\boldtheta}^{\otimes N}}[R(\boldf_{{\mathcal A(S)}})-R(\boldf_{\boldtheta})]=\Omega(\frac{T}{N}).
	\]
	In particular, when $\cA(S)=\hat{\boldtheta}_S$, there exists an instance of the Gaussian denoise problem where the expected prediction error of the optimal linear QML model on the training set $S$ is at least $\Omega(\frac{T}{N})$.
	\end{theorem}
	Theorem~\ref{thm:main_lower} implies that the optimal linear QML models on training sets achieve \emph{near-optimal prediction error}, indicating their utility in related QML tasks with theoretical guarantees.

	 \begin{figure*}[htp]
		\centering
		\includegraphics[width=0.97\textwidth]{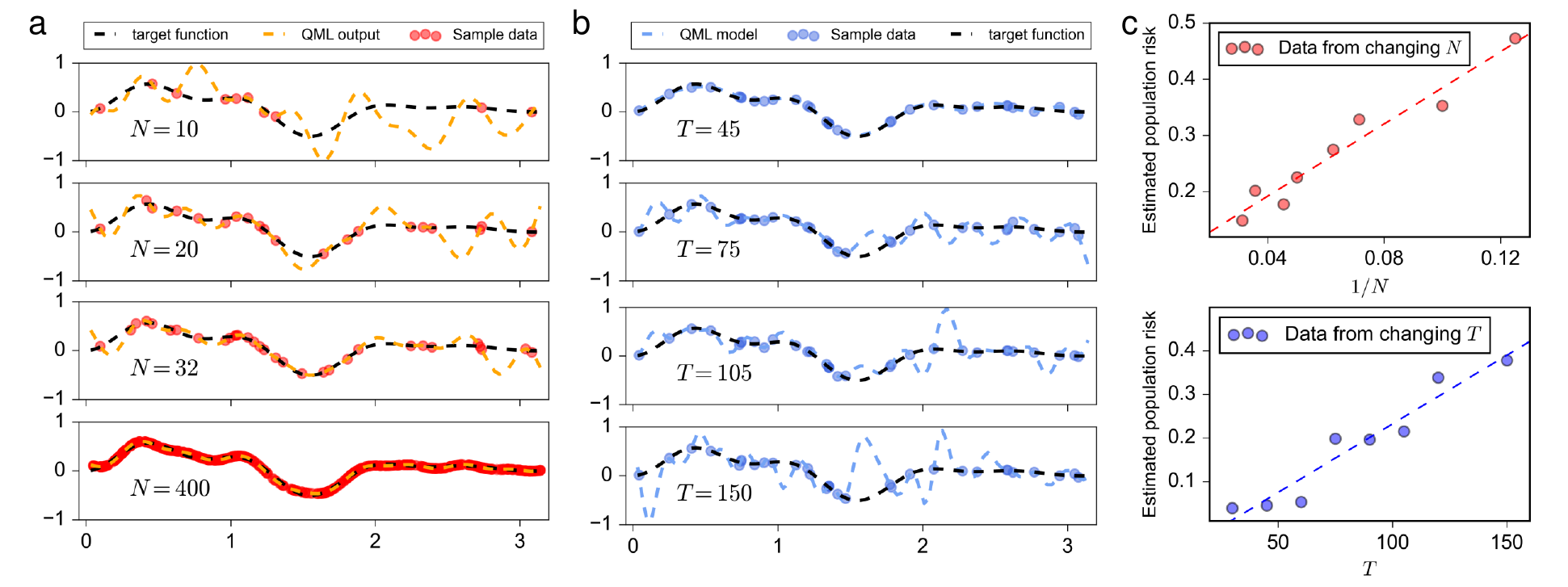}
		\caption{\small{\textbf{Curve fitting results about the prediction error.} Curve fitting is performed using a single-qubit parameterized quantum circuit. The QML model takes the $x$-label of the sample data as input, yielding an output within $[-1, 1]$. (a) The number of trainable parameters is fixed to be $60$. The fitting results under different training data sizes are demonstrated. (b) The training data size is fixed to be $32$. The fitting results under different numbers of trainable parameters are plotted. (c) The average loss of the optimal QML model on a training data set scales linearly in $1/N$ and $T$. In this task, the average loss of the optimal QML model is considered to be 0 when $T\geq 45$. Consequently, the average loss of the optimal QML model on a training dataset serves as a reliable indicator of its prediction error.
		}}
		\label{fig:experiment}
	\end{figure*}
	
	\begin{figure}[htp]
		\centering
		\includegraphics[width=0.47\textwidth]{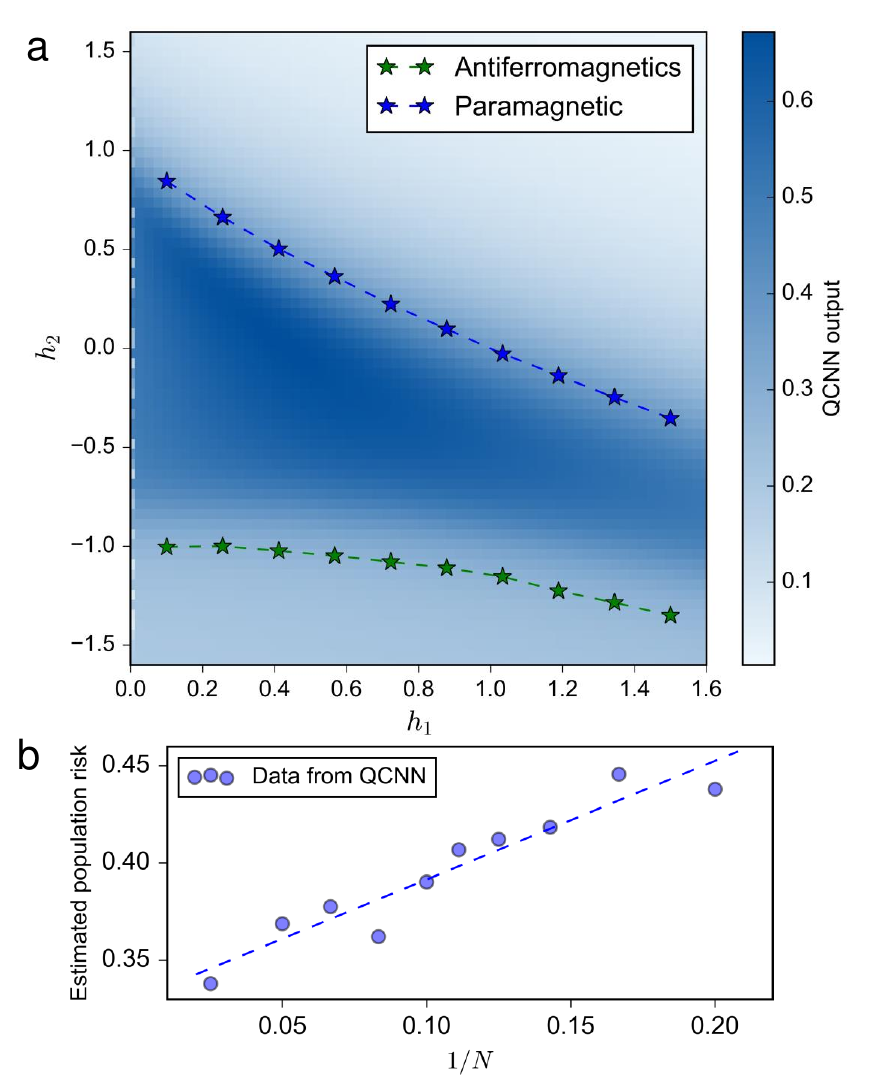}
		\caption{\small{\textbf{Classification results about the prediction error.} 
				(a) The phase diagram of the Hamiltonian. The phase boundary points (blue and red stars) are extracted from numerical simulations guided by \cite{congQuantumConvolutionalNeural2019}. The ground state indicates the paramagnetic phase if $(h_1, h_2)$ lies above the blue dotted curve. Conversely, the ground state exhibits the antiferromagnetic phase if it falls below the green dotted curve. The ground state indicates the SPT phase if it falls between the green and blue dotted curves. The background shading illustrates the output from the QCNN with $N=40$. (b) The average loss and prediction error of the optimal QML model on a training data set scales linearly in $1/N$.
		}}
		\label{fig:experiment2}
	\end{figure}
	
	\paragraph*{Numerical results.}
	We numerically simulate two QML tasks to support Theorem~\ref{thm:main_upper}, where we directly compute the parameterized quantum circuit executions by classical computations via the MindSpore Quantum framework~\cite{xu2024mindspore}.  
	
	We first consider learning a univariate analytic function using single-qubit data re-uploading QML model~\cite{NEURIPS2022_b250de41}. The advantages of simulating such a task are: (1) The target function belongs to the hypothesis spaces, and (2) single-qubit quantum circuits are easy to compute classically.    
	
	For instance, let the target function be 
	\[
	\boldf_{\boldtheta^*}(x)=\frac{\sin(3x)}{3x}-\frac{\sin(5x)}{5x}+\frac{\sin(7x)}{7x}-\frac{\sin(9x)}{9x}.
	\]
	We set the domain set be $\mathcal{X}=[0, \pi]$ and let the label set be $\mathcal{Y}=\{\boldf_{\boldtheta^*}(x):~x\in[0,\pi]\}$. The training data size varies from $N\in\{6, 8, 10, 12, 14, 16, 20, 24, 28, 32, 400\}$, obtained by taking i.i.d.~uniform points from $\mathcal X$. The number of trainable parameters varies from $T\in\{30, 45, 60, 75, 90, 105, 120, 150\}$. We use the \(\ell^2\) loss and approximate the optimal QML model on the given training set using Adam optimizer~\cite{KingBa15} and find $\boldf_{\boldtheta}\approx\boldf_{\hat\boldtheta}$ satisfying $\hat{R}_S(\boldf_{\boldtheta})\leq 0.001$. Specifically, the prediction error of a hypothesis function $\boldf_{\boldtheta}$ is approximated by computing the average loss on $2000$ points sampled uniformly from $\mathcal X$, whose accuracy is guaranteed by the aforementioned works about generalization. The simulation results under various training data sizes and parameter counts are depicted in FIG.~\ref{fig:experiment} (a) and (b), respectively. Figure~\ref{fig:experiment}(c) illustrates that the prediction error scales linearly with $1/N$ and $T$, supporting our theoretical findings. More detailed information like the QML model implementation and optimization process can be seen in Appendix~\ref{appendix:numerical_simulations}.

	The second task is quantum phase recognition, wherein a parameterized quantum circuit is utilized to classify an input quantum state into a specific quantum phase of matter. Following the setting outlined in previous works \cite{congQuantumConvolutionalNeural2019,Wu2023quantumphase}, the input states are ground states of a family of $n$-qubit Hamiltonians $H(h_1,h_2)$ parameterized by $h_1$ and $h_2$: 
	\begin{equation*}
        \begin{split}
        H(h_1,h_2)=-\sum_{i=1}^{n-2}(Z_iX_{i+1}Z_{i+2})-h_1\sum_{i=1}^nX_i-h_2\sum_{i=1}^{n-1}(X_iX_{i+1}),
        \end{split}
	\end{equation*}
	where $X_i$ ($Z_i$) represents the Pauli-X (Z) operator acting on the $i$-th qubit. The goal is to classify the input state into the symmetry-protected topological phases. The optimal classifier is illustrated by the two curves in FIG.~\ref{fig:experiment2}(a).
	
	To achieve this, we employ a quantum convolutional neural network (QCNN) \cite{congQuantumConvolutionalNeural2019} as our parameterized quantum circuit to classify the ground states of $\{H(h_1,h_2)\}$ on a 9-qubit system. The QCNN takes a quantum state as input and yields a real number in \([-1, 1]\) by measuring the expectation of the Pauli-X operator. The training set consists of $N$ ground states of $\{H(h_1,h_2);h_1\in [0, 1.6], h_2=0\}$ and their labels. We set $N=\{5, 6, 7, 8, 9, 10, 12, 15, 20, 40\}$ in our experiments. We use the \(\ell^2\) loss function and optimize the QCNN parameters by gradient descent, where the gradients are computed via the finite difference method. The optimization process terminates when the loss values exhibit minimal changes, yielding the QCNN model as an estimator of the optimal QML model on the training dataset \(\boldf_{\hat{\boldtheta}}\). Its prediction error is approximated by computing the average loss on $4096$ ground states of \(\{H(h_1,h_2);h_1\in [0, 1.6], h_2\in [-1.5, 1.5]\}\). 

	The simulation result for $N=40$ is depicted in FIG.~\ref{fig:experiment2} (a). Figure~\ref{fig:experiment2} (b) illustrates that the prediction error of the optimal QML model on a training data set scales linearly in $1/N$, which supports our theoretical findings. More detailed information and results regarding the QCNN model and these simulations are provided in Appendix~\ref{appendix:numerical_simulations}.

	\section{Discussion}
	In this work, our aim is to theoretically understand the performance of QML models caused by incomplete knowledge of the underlying distribution. Previous studies in QML literature focused on proving the generalization error bounds of QML models. This provides certain performance guarantees, but cannot explain the successful behaviors observed in recent numerical experiments, such as solving the quantum phase recognition problems. These findings highlight the need for a new analysis of the performance of QML models.
	
	We follow the theoretical framework proposed in~\cite{qiTheoreticalErrorPerformance2023,yu2024} and focus on the prediction error of the optimal QML models on a given training set. We prove that the prediction error scales linearly in $T/N$, quadratically improving the bound obtained from generalization error bounds. We also establish a lower bound for the prediction error. These theoretical findings are validated by numerical simulations. We shall mention that phenomena such as over-parameterization, double descent, or Barren Plateaus need to be considered when analyzing the optimization error, which will be investigated in future work to complete the theoretical analysis of QML models for solving supervised learning tasks.
	
	\section*{Methods}
	
	Our main technical contribution is a packing number lower bound for QML models that coincides with the covering number upper bound up to a constant factor, a result that may be of independent interest, as summarized in the following proposition.
	\begin{proposition}\label{prop_covering_packing}
		A family of linear QML models parameterized by $T$ trainable parameters is constructed as multiple copies of two-qubit amplitude-encoding circuits~\cite{perdomo2022preparation} and parameterized two-qubit unitaries~\cite{Vatan2004OptimalQuantumCircuit}, each acting on disjoint pairs of qubits with independent parameters.  
		The $\varepsilon$-covering entropy of this model class, with respect to the $L^{2}$ norm, scales as $O\!\left(T\log(1/\varepsilon)\right)$.  
		Moreover, its $\ubar{\varepsilon}$-packing entropy, also with respect to the $L^{2}$ norm, is lower bounded by $\Omega\!\left(T\log(1/\ubar{\varepsilon})\right)$.
	\end{proposition}
	The covering number upper bound is established using quantum-information-theoretic inequalities, following an approach similar to that of~\cite{du2022EfficientMeasureExpressivity,caroGeneralizationQuantumMachine2022}.  
	To prove the packing lower bound, we build on the results of~\citet{barthel2018fundamental} by designing a family of QML models whose hypothesis space is equivalent to a Grassmann manifold.
	This equivalence enables the application of established packing results for the Grassmann manifold~\cite{barthel2018fundamental} to compute the packing number of the QML models.
	A complete proof of Proposition~\ref{prop_covering_packing}, along with a detailed discussion of its implications, is provided in Appendix~\ref{app_covering_packing}.  
	This bound forms the cornerstone for deriving the upper and lower generalization guarantees in Theorems~\ref{thm:main_upper} and~\ref{thm:main_lower}.

	\paragraph*{Proof sketch of Theorem~\ref{thm:main_upper}.}
	To upper bound the prediction error, we first introduce an i.i.d.\ ghost sample set~\cite[Sec.~12]{probabilistic_theory_of_PR} and rewrite the error as a sum of random variables.  
	The expectation can be then upper bounded by integrating the corresponding tail probabilities. The later can be estimated by Bernstein's inequality after discretizing the hypothesis space using covering nets. In particular, the covering number of the hypothesis space of the data-reuploading QML models is estimated by converting the data-reuploading QML models into linear QML models~\cite{jerbi2023quantum}. A detailed proof and potential extensions can be found in Appendix~\ref{appen_thm1}. 
	
	\paragraph*{Proof sketch of Theorem~\ref{thm:main_lower}.}
	To lower bound the prediction error, we utilize an information-theoretic framework established by Yang and Barron~\cite{yang1999information} (see also~\cite[Sec.~15]{bach2021learning}). In particular, we discretize the hypothesis space using packing nets and reduce the problem into multiple hypothesis tests. Then, we lower bound the success probability using Fano's inequality. A detailed proof and discussion of Theorem~\ref{thm:main_lower} can be found in Appendix~\ref{appen_thm2}.

	\section*{Acknowledge}
	
	This work is supported by the National Key Research and Development Program of China (No.2024YFE0102500), the National Nature Science Foundation of China (No.\ 62302346, No.\ 12125103, No.\ 12071362, No.\ 12371424, No.\ 12371441), the Hubei Provincial Natural Science Foundation of China (No. 2024AFA045) and the ``Fundamental Research Funds for the Central Universities''. The first author Q. C. is sponsored by CPS-Yangtze Delta Region Industrial Innovation Center of Quantum and Information Technology-MindSpore Quantum Open Fund. The corresponding code for this project can be found at \url{https://github.com/qhc-nazgul/Near-optimal-Prediction-Error-Estimation-for-Quantum-Machine-Learning-Models.git}.

	\bibliography{main}
	
	\clearpage
	\newpage
	\appendix 
	\onecolumngrid
	
	\tableofcontents
	
	\medskip
	
	
\section{Preliminaries}
We unify some notations throughout the appendix. Let $[n]=\{1, 2, \cdots, n\}$. We use $U_N$ to denote the set of $N\times N$ unitary matrices.  The notation \(\cO\) is used for the measurement observable. The Bachmann-Landau symbols \(O\) and \(\Omega\) are used for upper and lower bounds. We use $\tilde{O}$ instead of $O$ when omitting the polylogarithmic dependence in the context. The logarithm function used throughout the appendix is assumed to have base 2.

We denote $\|A\|$ as the operator norm of a matrix $A$ and $\|v\|_2$ as the $L^2$ norm of a vector $v$. The $L^2(\mu)$ norm of a function $f(\bfx)$ with respect to the density $\mu(\bfx)$ is defined as \(\|\boldf\|_{L^2(\mu)}=\left(\int_{\mathcal{X}}|\boldf(\bfx)|^2 \mu(\bfx)d\bfx \right)^{1/2}\). When $\mu(\bfx)$ represents the density of the uniform distribution on the support of $\bfx$, $\|\cdot\|_{L^2(\mu)}$ is also denoted by $\|\cdot\|_{L^2}$. The $L^\infty$ norm of $\boldf$ is denoted and defined as $\|\boldf\|_{L^\infty}=\sup_\bfx|\boldf(\bfx)|$.
	
\subsection{Probability inequalities}
    We recall Markov inequality and Bernstein inequality, which play crucial roles in proving the main theorems. 
\begin{lemma}[Markov's inequality~{\cite[Prop.~1.2.4]{vershynin2018high}}]\label{lem_markov_ineq}
    For any non-negative random variable $X$ and $t>0$, we have 
    \begin{eqnarray}
        \mathbb{P}(X\geq t)\leq \frac{\mathbb{E}[X]}{t}. \nonumber
    \end{eqnarray}
\end{lemma}

\begin{lemma}[Bernstein's inequality for bounded random variables~{\cite[Thm.~2.8.4]{vershynin2018high}}]\label{thm:Bernstein_ineq_1}
    Let $X_1, X_2, \cdots, X_N$ be independent, mean zero random variables, such that $|X_i|\leq b$ for all $i\in[N]$. Then, for every $t\geq 0$, we have
    \begin{eqnarray}
        \bP(\bigl|\sum_{i=1}^{N}X_i\bigr|\geq t) \leq 2\exp(-\frac{t^2/2}{\sigma^2+bt}). \nonumber
    \end{eqnarray}
    Here $\sigma^2=\sum_{i=1}^{N}\bE [X_i^2]$ is the sum of variances. 
\end{lemma}
The following inequality can be easily obtained from Bernstein inequality and is more commonly used in learning theory.
\begin{corollary}\label{thm:Bernstein_ineq}
    Let $X_1, X_2, \cdots, X_N$ be independent random variables with uniformly bounded variances $\sigma^2$ and $\left|X-\mathbb{E}\left[X\right] \right|\leq b$. Then for $\forall t>0$ we have
    \begin{eqnarray}
        \mathbb{P}\left(\left|\mathbb{E}\left[X\right]-\frac{1}{N} \sum_{i=1}^N X_i\right| \geq t\right) \leq 2\exp \left(-\frac{N t^2}{2\left(\sigma^2+b t\right)}\right), \nonumber
    \end{eqnarray}
    which implies that \(\mathbb{P}\left(\mathbb{E}\left[X\right]-\frac{1}{N} \sum_{i=1}^N X_i \geq t\right) \leq \exp \left(-\frac{N t^2}{2\left(\sigma^2+b t\right)}\right)\).
\end{corollary}
\begin{proof}
    (of Corollary~\ref{thm:Bernstein_ineq})
    Lemma~\ref{thm:Bernstein_ineq_1} induces that
    \begin{equation}
    \begin{split}
    \mathbb{P}\left(\left|\mathbb{E}\left[X\right]-\frac{1}{N} \sum_{i=1}^N X_i\right| \geq t\right)=\mathbb{P}\left(\left|N\mathbb{E}\left[X\right]- \sum_{i=1}^N X_i\right| \geq Nt\right) &\leq 2\exp \left(-\frac{(N t)^2}{2\left(N\sigma^2+Nb t\right)}\right) \\
    &\leq 2\exp \left(-\frac{N t^2}{2\left(\sigma^2+b t\right)}\right), \nonumber
    \end{split}
    \end{equation}
     where \(X_1-\bE[X], X_2-\bE[X], \dots, X_N-\bE[X]\) are i.i.d. mean zero random variables with sum of variances $N\sigma^2$. 
\end{proof}

\begin{lemma}[Hoeffding's inequality for general bounded random variables~{\cite[Thm.~2.2.6]{vershynin2018high}}]\label{thm:Hoeffding_ineq_1}
    Let $X_1, \dots, X_N$ be independent random variables. Assume that $X_i\in[m_i, M_i]$ for $\forall i\in[N]$. Then, for any $t>0$, we have
    \[
    \bP(\sum_{i=1}^N(X_i-\bE X_i)\geq t)\leq \exp{-\frac{2t^2}{\sum_{i=1}^N(M_i-m_i)^2}}.
    \]
\end{lemma}

    \subsection{Packing and Covering}
The covering number, packing number and their corresponding metric entropies are defined as follows: 
\begin{definition}
    (Covering nets, covering numbers, covering entropy). Let $\cX$ be some set and $d$ be some measure that maps $\cX\times \cX \rightarrow \mathbb{R}$. Let $\cK \subset \cX$ be a subset and let $\varepsilon>0$.
    \begin{itemize}
        \item An $\varepsilon$-covering net of $\cK$ is a set $\mathcal{N} \subseteq \cK$ which satisfies that for $\forall x \in \cK$, $\exists y \in \mathcal{N}$ such that $d(x, y) \leq \varepsilon$. That is, $\mathcal{N} \subseteq \cK$ is an $\varepsilon$-covering net of $\cK$ if and only if $\cK$ can be covered by $\varepsilon$-balls around the elements in $\mathcal{N}$.
        \item The $\varepsilon$-covering number of $\cK$ is the smallest possible cardinality of the $\varepsilon$-covering net of $\cK$, denoted as $|\mathcal{N}(\cK, \varepsilon, d)|$. The $\varepsilon$-covering entropy of $\cK$ is the logarithm of its  $\varepsilon$-covering number.
    \end{itemize}
\end{definition}
\begin{definition}
    (Packing nets, packing numbers, packing entropy). Let $\cX$ be some set and $d$ be some measure that maps $\cX\times \cX \rightarrow \mathbb{R}$. Let $\cK \subset \cX$ be a subset and let $\varepsilon>0$.
    \begin{itemize}
        \item An $\varepsilon$-packing net of $\cK$ is a set $\mathcal{M} \subseteq \cK$ which satisfies that for $\forall x\neq y \in \mathcal{M},~d(x, y) > \varepsilon$. 
        \item The $\varepsilon$-packing number is the largest possible cardinality of the $\varepsilon$-packing net of $\cK$, denoted as $|\mathcal{M}(\cK, \varepsilon, d)|$. The $\varepsilon$-packing entropy (also known as the Kolmogorov capacity) of $\cK$ is the logarithm of its  $\varepsilon$-packing number.
    \end{itemize}
\end{definition}
These definitions are slight generalizations of the metric entropy notions introduced by~\citet{kolmogorov1959varepsilon}. When $(\cX, d)$ forms a metric space, the relationship between the covering entropy and packing entropy is shown by the following inequality:
\begin{eqnarray}\label{eqn:covering-and-packing}
    \log |\mathcal{M}(\cK, 2\varepsilon, d)| \leq \log |\mathcal{N}(\cK, \varepsilon, d)| \leq  \log |\mathcal{M}(\cK, \varepsilon, d)|.
\end{eqnarray}
Therefore, the covering and packing number (entropy) in metric spaces are of the same order. 

\subsection{Basic concepts from information theory}
We recall some basic concepts in information theory (cf.~\cite{ElementofInformationTheory}).
\paragraph{Entropy.} The (Shannon's) entropy of a random variable $X$ with density $p(x)$ $\forall x\in\mathcal{X}$ (with finite elements) is defined as: 
\[H(X)=H(p)=-\sum_{x\in\cX}p(x)\log p(x).\]
\paragraph{Kullback-Leibler divergence.} The Kullback-Leibler divergence or relative entropy between two densities $p(x)$ and $q(x) ~\forall x\in\mathcal{X}$ is defined as 
\[\KL(p||q)=\sum_{x\in\mathcal{X}}p(x)\log\frac{p(x)}{q(x)}. \]
Note that $\KL(p||q)$ is always nonnegative and is zero if $p(x)=q(x)~\forall x\in\mathcal{X}$.
\paragraph{Mutual Information.}\label{mutual_info} Consider two random variables $X$ and $Y$ with a joint density $p(x, y)$ and marginal density $p(x)$ and $p(y)$ for $\forall x\in\mathcal{X}$ and $y\in\mathcal{Y}$. The mutual information $I(X;~Y)$ is defined as:
\[
I(X;~Y)=\sum_{x\in\mathcal{X}}\sum_{y\in\mathcal{Y}} p(x, y)\log \frac{p(x, y)}{p(x)p(y)}.
\]
\begin{theorem}[Fano's inequality (cf.~Thm~2.10.1 in~\cite{ElementofInformationTheory}]
\label{Fano_ineq} 
For random variables $X, Y,\hat{X}$, if $X\rightarrow Y\rightarrow \hat{X}$ be a Markov chain, we have
\[
\bP(\hat{X}\neq X)\geq\frac{H(X)-I(X;~Y)-\log 2}{\log|\mathcal{X}|}.
\]
\end{theorem}

	\section{Bounds for the covering and packing entropies of Linear QML models}\label{app_covering_packing}
	We prove explicit upper and lower bounds for the covering and packing numbers of the hypothesis spaces of linear QML models, respectively. These bounds will be important for proving the prediction error upper and lower bounds.
	
	We work with the linear QML model $(V(\bfx),U(\boldtheta),\cO)$, where
	\begin{itemize}
		\item $V(\bfx)$ is an encoding mapping from each classical data vector $\bfx\in\mathcal{X}$ to a parametrized quantum circuit $V(\bfx)$ with a fixed circuit structure.
		\item $U(\boldtheta)$ is an training mapping from each parameter configuration $\boldtheta\in\bm{\Theta}$ to a parametrized quantum circuit $U(\boldtheta)$ with a fixed circuit structure.
		\item $\cO$ is a pre-fixed measurement observable with bounded operator norm for extracting classical information from the QML model.
	\end{itemize}
	In this way, the hypothesis function of the linear QML model $(V(\bfx),U(\boldtheta),\cO)$ with respect to a parameter configuration $\boldtheta$ is given by $\boldf_{\boldtheta}(\bfx)=\bra{0^n}V^\dagger(\bfx)U({\boldtheta})^\dagger\cO U({\boldtheta})V(\bfx)\ket{0^n}$. We define the hypothesis space $\cH=\{\boldf_{\boldtheta}:~\boldtheta\in\bm\Theta\}$. 
	
	\subsection{Covering entropy upper bound}
	We first establish a covering number upper bound for $\cH$ with respect to the $L^2(\mu)$ norm and absolute norm. Note that the following covering number upper bound of the set $\mathcal{H}_{\cO}:=\{U(\boldtheta)^{\dagger}\cO U(\boldtheta): \boldtheta\in\bm{\Theta} \}$ was established in~\cite{PhysRevLett.128.080506}:
	\begin{lemma}[Lemma 2 in~\cite{PhysRevLett.128.080506}]\label{thm:covering_number_circuit} 
		Let $U(\boldtheta)$ be a parametrized quantum circuit composed of $T$ parameterized 2-qubit quantum gates and an arbitrary number of non-trainable quantum gates. Let $\mathcal{H}_{\cO}:=\{U(\boldtheta)^{\dagger}\cO U(\boldtheta): \boldtheta\in\bm{\Theta} \}$. For $\varepsilon\in[0, 0.1)$, the $\varepsilon$-covering entropy for the hypothesis space $\mathcal{H}_{\cO}$ with respect to the operator norm $\|\cdot\|$ satisfies 
		\begin{eqnarray}\label{eqn:B3}
			\log \left(\left|\mathcal{N}(\mathcal{H}_{\cO}, \varepsilon, \|\cdot\|)\right|\right) \leq 16T\log(\frac{7T\|\cO\|}{\varepsilon}).
		\end{eqnarray}
	\end{lemma}
	We apply Lemma~\ref{thm:covering_number_circuit} to prove the following:
	\begin{theorem}[Covering entropy upper bound for $\cH$ with $L^2(\mu)$ norm]\label{lemma:covering-number-L2norm}
		Let $(V(\bfx), U(\boldtheta),\cO)$ be a linear quantum model where $V(\bfx)$ is an arbitrary encoding mapping, $U(\boldtheta)$ is a training mapping acting on parameter vectors of length $T$, and $\cO$ is a measurement observable with bounded operator norm. Then for any $\varepsilon\in[0, 0.1)$ and any density $\mu$, the $\varepsilon$-covering entropy of $\mathcal{H}$ with respect to $L^2(\mu)$ norm $\|\cdot\|_{L^2(\mu)}$ satisfies
		\begin{eqnarray}\label{appen_covering_upper_bound111}
			\log \left(\left|\mathcal{N}(\mathcal{H}, \varepsilon, \|\cdot\|_{L^2(\mu)})\right|\right)\leq \log \left(\left|\mathcal{N}(\mathcal{H}_{\cO}, \varepsilon, \|\cdot\|)\right|\right) \leq 16T\log(\frac{7T\|\cO\|}{\varepsilon}).
		\end{eqnarray}
	\end{theorem}
	\begin{proof}
		From Lemma~\ref{thm:covering_number_circuit} we know that for any $U(\boldtheta)^{\dagger}\cO U(\boldtheta)\in\mathcal{H}_{\cO}$, there exists $U(\boldtheta_{\varepsilon})^{\dagger}\cO U(\boldtheta_{\varepsilon})\in\mathcal{N}(\mathcal{H}_{\cO}, \varepsilon, \|\cdot\|)$, such that $\|U(\boldtheta_{\varepsilon})^{\dagger}\cO U(\boldtheta_{\varepsilon}) - U(\boldtheta)^{\dagger}\cO U(\boldtheta) \|\leq\varepsilon$. Let 
		\[
		\mathcal{N}=\{\boldf_{{\boldtheta}_{\varepsilon}}(\bfx)=\bra{0^n}V^\dagger(\bfx)U(\boldtheta_{\varepsilon})^\dagger\cO U(\boldtheta_{\varepsilon})V(\bfx)\ket{0^n}:~U(\boldtheta_{\varepsilon})^{\dagger}\cO U(\boldtheta_{\varepsilon})\in\mathcal{N}(\mathcal{H}_{\cO}, \varepsilon, \|\cdot\|)\}.
		\]
		Note that for any $\boldf_{\boldtheta}\in\cH$, its $L^2(\mu)$ distance with $\boldf_{{\boldtheta}_{\varepsilon}}$ is upper bounded by
		\begin{eqnarray}\label{appen_eqn_covering_number1}
			& & \left\| \bra{0^n}V^\dagger(\bfx)U({\boldtheta})^\dagger\cO U({\boldtheta})V(\bfx)\ket{0^n}-\bra{0^n}V^\dagger(\bfx)U(\boldtheta_{\varepsilon})^\dagger\cO U(\boldtheta_{\varepsilon})V(\bfx)\ket{0^n} \right\|_{L^2(\mu)} \nonumber \\
			&=& \left(\int_{\mathcal{X}} \left| \bra{0^n}V^\dagger(\bfx)U({\boldtheta})^\dagger\cO U({\boldtheta})V(\bfx)\ket{0^n}-\bra{0^n}V^\dagger(\bfx)U(\boldtheta_{\varepsilon})^\dagger\cO U(\boldtheta_{\varepsilon})V(\bfx)\ket{0^n} \right|^2 \mu(\bfx) d\bfx   \right)^{1/2} \nonumber \\
			&=& \left(\int_{\mathcal{X}} \left| \Tr\left(U({\boldtheta})^\dagger\cO U({\boldtheta})V(\bfx)\ket{0^n}\bra{0^n}V^\dagger(\bfx)\right) - \Tr\left(U(\boldtheta_{\varepsilon})^\dagger\cO U(\boldtheta_{\varepsilon})V(\bfx)\ket{0^n}\bra{0^n}V^\dagger(\bfx)\right) \right|^2 \mu(\bfx) d\bfx  \right)^{1/2} \nonumber \\
			&=& \left(\int_{\mathcal{X}} \left|\Tr((U({\boldtheta})^\dagger\cO U({\boldtheta})-U(\boldtheta_{\varepsilon})^\dagger\cO U(\boldtheta_{\varepsilon}))V(\bfx)\ket{0^n}\bra{0^n}V^\dagger(\bfx))\right|^2 \mu(\bfx) d\bfx \right)^{1/2} \nonumber \\
			&\leq& \left(\|U(\boldtheta)^{\dagger}\cO U(\boldtheta) - U(\boldtheta_{\varepsilon})^{\dagger}\cO U(\boldtheta_{\varepsilon}) \|^2 \int_{\mathcal{X}} \Tr[V(\bfx)\ket{0^n}\bra{0^n}V^\dagger(\bfx)]^2  \mu(\bfx) d\bfx \right) ^{1/2} \nonumber \\
			&=& \|U(\boldtheta)^{\dagger}\cO U(\boldtheta) - U(\boldtheta_{\varepsilon})^{\dagger}\cO U(\boldtheta_{\varepsilon}) \| \leq\varepsilon, 
		\end{eqnarray}
            where the inequality holds since $\|A\|=\max_{\ket{\psi}}\bra{\psi}A\ket{\psi}$ and the last equality holds since $\mu$ is a probability density.
		Thus, $\mathcal{N}$ is an $\varepsilon$-covering net of $\cH$ with respect to the $L^2(\mu)$ norm, which implies that
		\[
		\log \left(\left|\mathcal{N}(\mathcal{H}, \varepsilon, \|\cdot\|_{L^2(\mu)})\right|\right)\leq \log \left(\left|\mathcal{N}(\mathcal{H}_{\cO}, \varepsilon, \|\cdot\|)\right|\right) \leq 16T\log(\frac{7T\|\cO\|}{\varepsilon}).
		\]
		Note that this inequality holds for any density $\mu$.
	\end{proof}
	We can prove a similar covering number bound for $\cH$ under the absolute norm $|\cdot|$:
	\begin{corollary}\label{corollary:covering-number-abs-norm}
		Under the same conditions of Theorem~\ref{lemma:covering-number-L2norm}, we have that 
		\begin{eqnarray}
			\log \left(\left|\mathcal{N}(\mathcal{H}, \varepsilon, |\cdot|)\right|\right) \leq 16T\log(\frac{7T\|\cO\|}{\varepsilon}).
		\end{eqnarray}
	\end{corollary}
	\begin{proof}
		Note that 
		\begin{eqnarray}
			\left| \boldf_{{\boldtheta}}(\bfx) - \boldf_{{\boldtheta}_{\varepsilon}}(\bfx) \right| 
			&=& | \bra{0^n}V^\dagger(\bfx)U({\boldtheta})^\dagger\cO U({\boldtheta})V(\bfx)\ket{0^n}-\bra{0^n}V^\dagger(\bfx)U(\boldtheta_{\varepsilon})^\dagger\cO U(\boldtheta_{\varepsilon})V(\bfx)\ket{0^n} | \nonumber \\
			&=& |\bra{0^n}V^\dagger(\bfx)(U({\boldtheta})^\dagger\cO U({\boldtheta})-U(\boldtheta_{\varepsilon})^\dagger\cO U(\boldtheta_{\varepsilon}))V(\bfx)\ket{0^n}| \nonumber \\
			&\leq& \|U(\boldtheta)^{\dagger}\cO  U(\boldtheta) - U(\boldtheta_{\varepsilon})^{\dagger}\cO  U(\boldtheta_{\varepsilon}) \|.  \nonumber
		\end{eqnarray}
		The proof goes similarly as the proof of~\Cref{lemma:covering-number-L2norm}. 
	\end{proof}
	
	\subsection{Packing entropy lower bound}\label{packing_lower_bound}
	
\begin{figure}[ht!]
		\centering
		\includegraphics[width=0.8\textwidth]{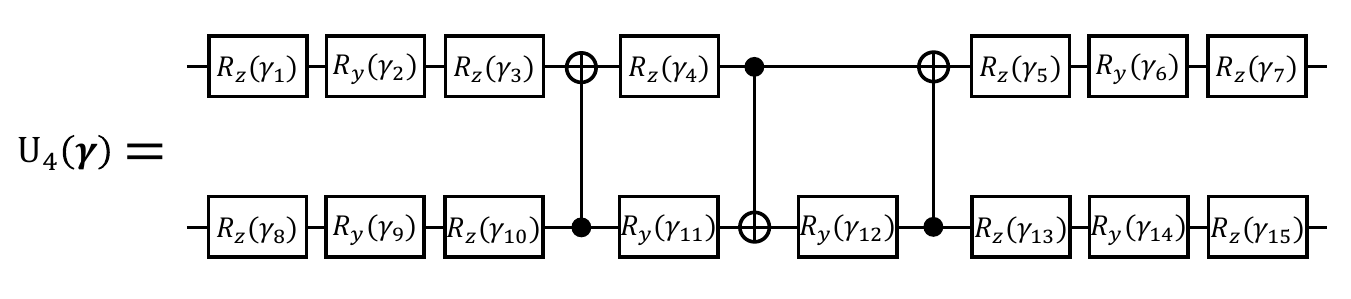}
		\caption{\small{\textbf{A parameterized quantum circuit to represent $U_4$.} Note that $\{\text{U}_4(\boldsymbol{\gamma}):\boldsymbol{\gamma}\in(0,2\pi]^{15}\}\cong U_4$. }}
		\label{fig:universal_U4}
	\end{figure}

\begin{figure}[ht!]
		\centering
		\includegraphics[width=0.3\textwidth]{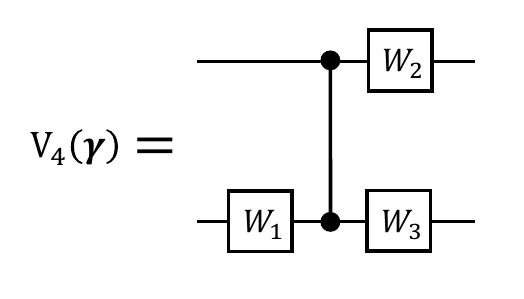}
		\caption{\small{\textbf{A parameterized quantum circuit for two-qubit amplitude encoding.} For any unit vector $\boldsymbol{\gamma}\in\mathbb{C}^4$, let $\ket{\boldsymbol{\gamma}}=\gamma_{0}\ket{00}+\gamma_1\ket{01}+\gamma_2\ket{10}+\gamma_3\ket{11}$. It holds that $\ket{\boldsymbol{\gamma}}=\text{V}_4(\boldsymbol{\gamma})^{\dagger}\ket{00}$. We specify the choice of $W_1,W_2,W_3\in U_2$: Let $\texttt{Unitary(a, b)}={1}/{\sqrt{|\texttt{a}|^2+|\texttt{b}|^2}}\left(\begin{smallmatrix}\texttt{a}&\texttt{b}\\-\texttt{b}^*&\texttt{a}^*\end{smallmatrix}\right)$ be a $2\times 2$ unitary matrix encoded with parameters $\texttt{a}$ and $\texttt{b}$. For any unit vector $\boldsymbol{\gamma}\in\mathbb{C}^4$, set $A_1=[\gamma_0\phantom{0}\gamma_1]^T$ and $A_2=[\gamma_2\phantom{0}\gamma_3]^T$. If $A_1^{\dagger}A_2=0$, let $k={\|A_2\|_2}/{\|A_1\|_2}$; otherwise, let $k=-{A_1^{\dagger}A_2}/{|A_1^{\dagger}A_2|}\cdot{\|A_2\|_2}/{\|A_1\|_2}$. Set $W_1=\texttt{Unitary}(\gamma_3-k\gamma_1, \gamma_2^*-k\gamma_0^*)^T$. Let $\text{CZ}(\mathbb{I}\otimes W_1)\ket{\boldsymbol{\gamma}}=\epsilon_0\ket{00}+\epsilon_1\ket{01}+\epsilon_2\ket{10}+\epsilon_3\ket{11}$ then $W_2=\texttt{Unitary}(\epsilon_1^*, \epsilon_3^*)$. Let $(W_2\otimes \mathbb{I})\text{CZ}(\mathbb{I}\otimes W_1)\ket{\boldsymbol{\gamma}}=\varrho_0\ket{00}+\varrho_1\ket{01}+\varrho_2\ket{10}+\varrho_3\ket{11}$ then $W_3=\texttt{Unitary}(\varrho_0^*, \varrho_1^*)^T$.
		}}
		\label{fig:amplitude_encoding}
	\end{figure}

	We work with a restricted linear QML model named $2$-local linear QML model, i.e.~all data encoding and training mappings are formed by parameterized quantum circuits acting locally on $2$ nonoverlapping qubits. We use the parameterized quantum circuit $\text{U}_4(\cdot)$ introduced in~\cite[Thm~5]{Vatan2004OptimalQuantumCircuit} to read and proceed parameters (see FIG.~\ref{fig:universal_U4}). For data encoding, we utilize the amplitude encoding circuit $\text{V}_4^\dagger(\cdot)$ introduced in~\cite[Prop~3.2]{perdomo2022preparation} to encode any $4$-dimensional unit vectors into the amplitude of 2-qubit state (see FIG.~\ref{fig:amplitude_encoding}). 
	
\begin{figure}[ht!]
		\centering
		\includegraphics[width=0.39\textwidth]{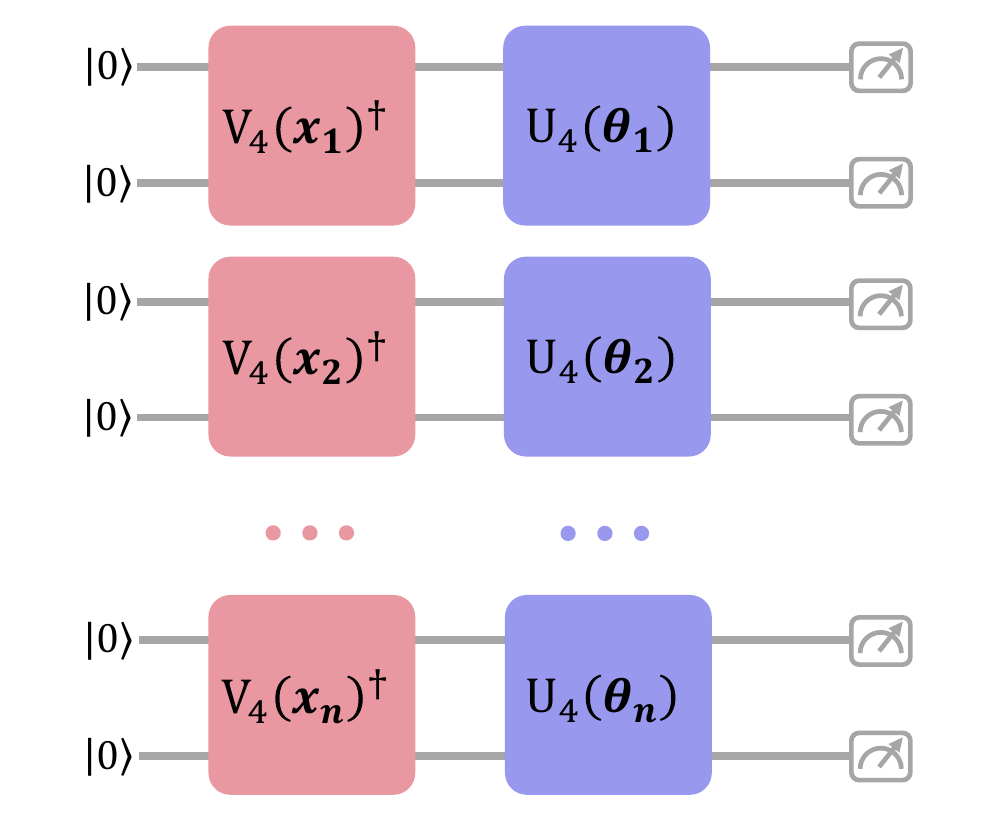}
		\caption{\small{\textbf{Template circuit to represent \(\cH_{2-local}\).} Every angle from the single-qubit rotation gates comes from \((0, 2\pi]\). 
		}}
		\label{fig:universal_GDP}
	\end{figure}

	We define the 2-local linear QML model with the following restrictions:
	\begin{itemize}
            \item The (unknown) distribution (density) $\mu$ over $\bfx\in\cX$ is uniformly bounded from below; i.e.~there exists a constant $0< C_1< 1$ such that $\mu(\bfx)\geq C_1~\forall \bfx\in\mathcal{X}$.
		\item The data encoding mapping $V(\bfx)$ encodes a partially normalized classical data vector $\bfx=[\mathbf{x}_1,\dots,\mathbf{x}_{n}]\in\mathbb{C}^{4n}$, where $\mathbf{x}_i\in\mathbb{C}^4$ is a unit vector for $\forall i\in[n]$, into a $2n$-qubit quantum circuit $V(\bfx)=\text{V}_4^{\dagger}(\mathbf{x}_1)\otimes \cdots\otimes\text{V}_4^{\dagger}(\mathbf{x}_n)$. 
		\item The training mapping encodes a parameter configuration $\boldtheta=[\boldtheta_1,\dots,\boldtheta_n]\in\mathbf{\Theta}_{2-local}=(0,2\pi]^{15n}$ into a $2n$-qubit quantum circuit $U(\boldtheta)=\text{U}_4(\boldtheta_1)\otimes\cdots\otimes\text{U}_4(\boldtheta_n)$. 
		\item The measurement observable is a sum of $2$-local measurements, i.e.~\(\cO=\alpha_1\ket{o_1}\bra{o_1}\otimes\bI_{2(n-1)}+\alpha_2\bI_2\otimes\ket{o_2}\bra{o_2}\otimes\bI_{2(n-2)} + \cdots + \alpha_n\bI_{2(n-1)}\otimes\ket{o_{n}}\bra{o_{n}}\), where $\alpha_i\in\mathbb{C}$, \(\ket{o_i}\in\mathbb{C}^4\) for all \(i\in[n]\) to ensure that $\|\cO\|$ is bounded. In particular, we can choose $\alpha_1,\dots,\alpha_n$ such that $\max_{i\in[n]}|\alpha_i|=\Omega(1)$.
	\end{itemize}
	In this setting, the hypothesis functions can be evaluated by the parameterized quantum circuit illustrated in FIG.~\ref{fig:universal_GDP}. Let the hypothesis space be denoted as 
	\[
	\cH_{2-local}=\{\boldf_{\boldtheta}(\bfx)=\bra{0^n}V^\dagger(\bfx)U({\boldtheta})^\dagger\cO U({\boldtheta})V(\bfx)\ket{0^n}:~\boldtheta\in(0,2\pi]^{15n}\}.
	\]
	Note that the number of parameters $T$ equals $15n$. Due to the $2$-local constraint, we can decompose each hypothesis function into the following form:
	\[
	\boldf_{\boldtheta}(\bfx)=\bra{0^{2n}}V^{\dagger}(x)U^{\dagger}(\boldtheta)\cO U(\boldtheta)V(\bfx)\ket{0^{2n}}=\sum_{i=1}^{n}\alpha_i\bra{00}\text{V}_4(\mathbf{x}_i) \text{U}_4(\boldtheta_i) ^{\dagger} \ket{o_i}\bra{o_i}\text{U}_4(\boldtheta_i) \text{V}_4(\mathbf{x}_i)^{\dagger}  \ket{00},
	\]
	where the normalized vector \(\mathbf{x}_i\in\mathbb{C}^4\), the parameter configuration \(\boldtheta_i\in(0,2\pi]^{15}\) for $\forall i\in[n]$. Let 
    $$\cH_i=\{\boldf_{\boldtheta_i}^{(i)}(\mathbf{x}_i)=\alpha_i\bra{00}\text{V}_4(\mathbf{x}_i) \text{U}_4(\boldtheta_i) ^{\dagger} \ket{o_i}\bra{o_i}\text{U}_4(\boldtheta_i) \text{V}_4(\mathbf{x}_i)^{\dagger}  \ket{00}:~\boldtheta_i\in(0,2\pi]^{15}\} \text{ for } \forall i\in[n].$$ 
    The hypothesis space $\cH_{2-local}$ satisfies that $\cH_{2-local}=\cH_1\oplus\cdots\oplus\cH_n$. 
	
	We first estimate the packing number of $\cH_i$. Note that for a fixed state $\ket{o_i}\in\mathbb{C}^4$, the set $\left\{\text{U}_4(\boldtheta_i) ^{\dagger} \ket{o_i}\bra{o_i}\text{U}_4(\boldtheta_i); \boldtheta_i\in(0, 2\pi]^{15}\right\}$ is precisely the Grassmannian \(\text{Gr}_{1, 4}\), i.e.~the set of all $4\times 4$ orthonormal projections. \citet{barthel2018fundamental} have established \(\varepsilon\)-packing number bounds of $\text{Gr}_{1,m}$ with respect to the operator norm: 
	\[\frac{1}{19m^2}\left(\frac{9}{5\varepsilon}\right)^{2(m-1)}\leq\left|\mathcal{M}\left(\text{Gr}_{1,m}, \varepsilon, \|\cdot\|\right)\right|\leq 38m^2\left(\frac{3}{2\varepsilon}\right)^{2(m-1)}. \]
	Therefore, for a given $i\in[n]$, there exists an \(\varepsilon\)-packing net \(\mathcal{M}^{(i)}=\{\ket{v_{i, j}}\!\bra{v_{i, j}}:~j\in[|\mathcal{M}^{(i)}|\}\) with respect to the operator norm \(\|\cdot\|\), where \(|\mathcal{M}^{(i)}|=\Theta({1}/{\varepsilon}^6)\), such that for any two distinct elements \(\ket{v_{i, j_1}}\!\bra{v_{i, j_1}}\) and \(\ket{v_{i, j_2}}\!\bra{v_{i, j_2}}\) in \(\mathcal{M}^{(i)}\), 
	\[
	\|\ket{v_{i, j_1}}\!\bra{v_{i, j_1}}-\ket{v_{i, j_2}}\!\bra{v_{i, j_2}}\|>\varepsilon.
	\] 
	We consider the set $\mathcal{M}_{\boldf}^{(i)}=\{\alpha_i\bra{00}\text{V}_4(\mathbf{x}_i)\ket{v_{i, j}}\!\bra{v_{i, j}}\text{V}_4(\mathbf{x}_i)^{\dagger} \ket{00}: \ket{v_{i, j}}\!\bra{v_{i, j}}\in\mathcal{M}^i\}\subseteq \cH_i$. In particular, for any two distinct elements \[\alpha_i\bra{00}\text{V}_4(\mathbf{x}_i)\ket{v_{i, j_1}}\!\bra{v_{i, j_1}}\text{V}_4(\mathbf{x}_i)^{\dagger} \ket{00}\] and \[\alpha_i\bra{00}\text{V}_4(\mathbf{x}_i)\ket{v_{i, j_2}}\!\bra{v_{i, j_2}}\text{V}_4(\mathbf{x}_i)^{\dagger} \ket{00}\] in $\mathcal{M}_{\boldf}^{(i)}$, we consider their $L^2(\mu)$ distance 
	\begin{eqnarray}
		~ &~& \|\alpha_i\bra{00}\text{V}_4(\mathbf{x}_i)\ket{v_{i, j_1}}\!\bra{v_{i, j_1}}\text{V}_4(\mathbf{x}_i)^{\dagger} \ket{00}-\alpha_i\bra{00}\text{V}_4(\mathbf{x}_i)\ket{v_{i, j_2}}\!\bra{v_{i, j_2}}\text{V}_4(\mathbf{x}_i)^{\dagger} \ket{00}\|_{L^2(\mu)} \nonumber \\
		&=&|\alpha_i| \left(\int_{\mathbf{x}_i\in\mathbb{C}^4, \|\mathbf{x}_i\|=1}\left(\bra{00}\text{V}_4(\mathbf{x}_i)\ket{v_{i, j_1}}\!\bra{v_{i, j_1}}\text{V}_4(\mathbf{x}_i)^{\dagger} \ket{00}- \right. \right. \nonumber \\
		&& \qquad \qquad \qquad\qquad\qquad\qquad\qquad\qquad\qquad \left. \left. \bra{00}\text{V}_4(\mathbf{x}_i)\ket{v_{i, j_2}}\!\bra{v_{i, j_2}}\text{V}_4(\mathbf{x}_i)^{\dagger} \ket{00}\right)^2 \mu(\mathbf{x}_i)d\bfx_i \right)^{1/2} \nonumber \\
		&\geq& C_1|\alpha_i| \left(\int_{\mathbf{x}_i\in\mathbb{C}^4, \|\mathbf{x}_i\|=1}\left(\bra{00}\text{V}_4(\mathbf{x}_i)\ket{v_{i, j_1}}\!\bra{v_{i, j_1}}\text{V}_4(\mathbf{x}_i)^{\dagger} \ket{00}- \right. \right. \nonumber \\
		&& \qquad \qquad \qquad\qquad\qquad\qquad\qquad\qquad\qquad \left. \left. \bra{00}\text{V}_4(\mathbf{x}_i)\ket{v_{i, j_2}}\!\bra{v_{i, j_2}}\text{V}_4(\mathbf{x}_i)^{\dagger} \ket{00}\right)^2 d\mathbf{x}_i\right)^{1/2} \nonumber \\
		&\geq& C_1C_2|\alpha_i| \sup_{\mathbf{x}_i\in\mathbb{C}^4, \|\mathbf{x}_i\|=1} \left|\bra{00}\text{V}_4(\mathbf{x}_i)\ket{v_{i, j_1}}\!\bra{v_{i, j_1}}\text{V}_4(\mathbf{x}_i)^{\dagger} \ket{00}-\bra{00}\text{V}_4(\mathbf{x}_i)^{\dagger}\ket{v_{i, j_2}}\!\bra{v_{i, j_2}}\text{V}_4(\mathbf{x}_i)^{\dagger} \ket{00}\right| \nonumber \\
		&=&C_1C_2|\alpha_i| \|\ket{v_{i, j_1}}\!\bra{v_{i, j_1}}-\ket{v_{i, j_2}}\!\bra{v_{i, j_2}}\|>C_1C_2|\alpha_i|\varepsilon, \nonumber
	\end{eqnarray}
	where the first inequality uses the fact that the density $\mu$ is uniformly bounded from below, and the second inequality holds since there exists a constant $C_2$ such that \(\|f\|_{L^2}\geq C_2\|f\|_{L^\infty}\) for any polynomial function $f$~\cite[Lemma~4.5.3]{brenner2008MathematicalTheoryFinite}. Thus, $\mathcal{M}_{\boldf}^{(i)}$ is an $C_1C_2|\alpha_i|\varepsilon$-packing net of $\cH_i$, which implies that $|\mathcal{M}(\cH_i,\varepsilon,\|\cdot\|_{L^2(\mu)})|=\Omega(1/\varepsilon^6)$.
	
	To estimate $|\mathcal{M}(\cH_{2-local},\varepsilon,\|\cdot\|_{L^2(\mu)})|$, we construct the set $\mathcal{M}_{\boldf}=\mathcal{M}_{\boldf}^{(1)}\oplus\cdots\oplus\mathcal{M}_{\boldf}^{(n)}$. For two distinct elements \[\sum_{i=1}^{n}\alpha_i\bra{00}\text{V}_4(\mathbf{x}_i)\ket{v_{i, j_{i, 1}}}\!\bra{v_{i, j_{i, 1}}}\text{V}_4(\mathbf{x}_i)^{\dagger} \ket{00}\] and \[\sum_{i=1}^{n}\alpha_i\bra{00}\text{V}_4(\mathbf{x}_i)\ket{v_{i, j_{i, 2}}}\!\bra{v_{i, j_{i, 2}}}\text{V}_4(\mathbf{x}_i)^{\dagger} \ket{00},\] we have similarly that
	\begin{eqnarray}
		~ &~& \|\sum_{i=1}^n\alpha_i\bra{00}\text{V}_4(\mathbf{x}_i)\ket{v_{i, j_{i,1}}}\!\bra{v_{i, j_{i,1}}}\text{V}_4(\mathbf{x}_i)^{\dagger} \ket{00}- \nonumber \\ &~&\qquad\qquad\qquad\qquad\qquad\qquad\qquad\qquad \sum_{i=1}^n\alpha_i\bra{00}\text{V}_4(\mathbf{x}_i)\ket{v_{i, j_{i,2}}}\!\bra{v_{i, j_{i,2}}}\text{V}_4(\mathbf{x}_i)^{\dagger} \ket{00}\|_{L^2(\mu)} \nonumber \\
		&\geq&C_1\left(\int_{\mathbf{x}}\left|\sum_{i=1}^n\alpha_i\bra{00}\text{V}_4(\mathbf{x}_i)(\ket{v_{i, j_{i,1}}}\!\bra{v_{i, j_{i,1}}}-\ket{v_{i, j_{i,2}}}\!\bra{v_{i, j_{i,2}}})\text{V}_4(\mathbf{x}_i)^{\dagger} \ket{00}\right|^2 d\mathbf{x}\right)^{1/2} \nonumber \\
		&\geq&C_1C_2\sup_{\|\mathbf{x_i}\|=1, \forall i\in[n]} |\sum_{i=1}^n\alpha_i\bra{00}\text{V}_4(\mathbf{x}_i)(\ket{v_{i, j_{i,1}}}\!\bra{v_{i, j_{i,1}}}-\ket{v_{i, j_{i,2}}}\!\bra{v_{i, j_{i,2}}})\text{V}_4(\mathbf{x}_i)^{\dagger} \ket{00}| \nonumber \\
		&\geq&C_1C_2|\alpha_i| \|\ket{v_{i, j_{i,1}}}\!\bra{v_{i, j_{i,1}}}-\ket{v_{i, j_{i,2}}}\!\bra{v_{i, j_{i,2}}}\|>C_1C_2|\alpha_i|\varepsilon, \nonumber
	\end{eqnarray}
	where the last inequality holds for any $i\in[n]$, since we can choose $\bfx_i$ which maximize \[\bra{00}\text{V}_4(\mathbf{x}_i)(\ket{v_{i, j_{i,1}}}\!\bra{v_{i, j_{i,1}}}-\ket{v_{i, j_{i,2}}}\!\bra{v_{i, j_{i,2}}})\text{V}_4(\mathbf{x}_i)^{\dagger} \ket{00},\] and choose the other $\bfx_{i'}$ such that \[\bra{00}\text{V}_4(\mathbf{x}_{i'})(\ket{v_{{i'}, j_{{i'},1}}}\!\bra{v_{{i'}, j_{{i'},1}}}-\ket{v_{{i'}, j_{{i'},2}}}\!\bra{v_{{i'}, j_{{i'},2}}})\text{V}_4(\mathbf{x}_{i'})^{\dagger} \ket{00}=0.\] Summarizing the above, we have the following:
	\begin{theorem}[Packing entropy lower bound for $\cH_{2-local}$ with $L^2(\mu)$ norm]
		Let $(V(\bfx),U(\boldtheta),\cO)$ be a $2$-local linear quantum model with $T$ trainable parameters, where $V(\bfx)$ and $U{(\boldtheta)}$ are tensor products of the parameterized quantum circuit $\mathrm{V}_4$ and $\mathrm{U}_4$ respectively. The observable $\cO$ is a sum of $2$-local measurements with bounded operator norm. Suppose the density $\mu$ satisfies $\mu(\bfx)\geq C_1$ for any $\bfx\in\mathcal{X}$ and $C_1\in(0, 1)$. Then for any $\varepsilon\in [0, 0.014)$, the $\varepsilon$-packing entropy of $\mathcal{H}_{2-local}$ with respect to  $\|\cdot\|_{L^2(\mu)}$ satisfies
		\begin{eqnarray}\label{appen_packing_lower_bound111}
			\log|\mathcal{M}(\cH_{2-local},\varepsilon,\|\cdot\|_{L^2(\mu)})|=\Omega\left(T\log(\frac{1}{\varepsilon})\right).
		\end{eqnarray}
	\end{theorem}
    Since $\cH_{2-local}\subseteq\cH$, we have the following corollary:
    \begin{corollary}[Packing number lower bound for $\cH$ with $L^2(\mu)$ norm]\label{lemma:packing-number-L2norm}
    Let $(V(\bfx), U(\boldtheta),\cO)$ be a linear quantum model where $V(\bfx)$ is an arbitrary encoding mapping, $U(\boldtheta)$ is a training mapping acting on parameter vectors of length $T$, and $\cO$ is a measurement observable with bounded operator norm. 
    Suppose the density $\mu$ satisfies $\mu(\bfx)\geq C_1$ for any $\bfx\in\mathcal{X}$ and $C_1\in(0, 1)$. Then for any $\varepsilon\in [0, 0.014)$, the $\varepsilon$-packing entropy of $\mathcal{H}$ with respect to  $\|\cdot\|_{L^2(\mu)}$ satisfies
	\begin{eqnarray}\label{appen_packing_lower_bound1111}
			\log|\mathcal{M}(\cH,\varepsilon,\|\cdot\|_{L^2(\mu)})|=\Omega\left(T\log(\frac{1}{\varepsilon})\right).
		\end{eqnarray}
    \end{corollary}

	\section{Proof of Theorem~1}\label{appen_thm1}
	We proceed to prove~\Cref{thm:main_upper}, where we focus on supervised learning tasks characterized by an input set $\mathcal{X}$, an output (label) set $\mathcal{Y}$, and an underlying density $\mathcal{P}$ over $\mathcal{X}\times \mathcal{Y}$. We shall work with the loss function $\ell(\boldf_{\boldtheta}; \bfx, y):= (\boldf_{\boldtheta}(\bfx)-y)^2$. In this case, the best linear QML model in the hypothesis space $\cH$ is the prediction error minimizer $\boldf_{{\boldtheta}^{*}}$, where 
	\[{\boldtheta^*}\in\mathop{\arg\min}_{\boldtheta\in \bm{\Theta}} R(\boldf_{\boldtheta})=\mathop{\arg\min}_{\boldtheta\in \bm{\Theta}} \mathbb{E}_{(\bfx, y)\sim \cP}\left[ \ell(\boldf_{\boldtheta}; \bfx, y) \right].
	\] 
	Since the density $\mathcal{P}$ is usually unknown, it might be impossible to evaluate the objective function $R(\boldf_{\boldtheta})$. Instead, a finite training set $S=\{(\bfx_i, y_i)\in\mathcal{X}\times \mathcal{Y}:~i\in[N]\}$ of size $N$ is provided and one may computing the $\hat{R}_S(\boldf_{\boldtheta}) = \frac{1}{N}\sum_{i=1}^{N}\ell(\boldf_{\boldtheta}; \bfx_i, y_i)$ as an estimation of the prediction error $R(\boldf_{\boldtheta})$. QML algorithms/methods aims to find the optimal QML model on the training datset $\boldf_{\hat{\boldtheta}_S}$, where
	\[\hat{\boldtheta}_S\in\mathop{\arg\min}_{\boldtheta\in \bm{\Theta}} \hat{R}_S(\boldf_{\boldtheta})=\mathop{\arg\min}_{\boldtheta\in \bm{\Theta}} \frac{1}{N}\sum_{i=1}^{N}\ell(\boldf_{\boldtheta}; \bfx_i, y_i).
	\] 
	We shall ignore the optimization error incurred by specific optimization algorithms; our goal is to bound the differences between the empirical and prediction error minimizers in terms of the necessary sample size (sample complexity) and measures of the hypothesis space (model complexity). 
	
	For this purpose, the difference can be naturally measured by the prediction error: $\cE_P(\boldf_{\hat{\boldtheta}_S})=R(\boldf_{\hat{\boldtheta}_S})-R(\boldf_{{\boldtheta}^{*}})$, which is a random variable depending on the unknown density $\cP^{\otimes N}$ over the training data set $S$ of size $N$. We shall focus on estimating the expected value of the prediction error. Note that
	\begin{eqnarray}\label{eqn:sta_err}
		\mathbb{E}_{\cP^{\otimes N}}\left[\cE_P(\boldf_{\hat{\boldtheta}_S}) \right] &=& \mathbb{E}_{\cP^{\otimes N}}\left[ R(\boldf_{\hat{\boldtheta}_S})-R(\boldf_{\boldtheta^*}) \right] \nonumber \\
		&=& \mathbb{E}_{\cPN}\left[ R(\boldf_{\boldtheta^*})-2\hat{R}_{S}(\boldf_{\hat{\boldtheta}_S})+R(\boldf_{\hat{\boldtheta}_S}) \right] + 2\mathbb{E}_{\cPN}\left[ \hat{R}_{S}(\boldf_{\hat{\boldtheta}_S})- R(\boldf_{\boldtheta^*})\right] \nonumber \\
		~ &\leq& \mathbb{E}_{\cPN}\left[ R(\boldf_{\boldtheta^*})-2\hat{R}_{S}(\boldf_{\hat{\boldtheta}_S})+R(\boldf_{\hat{\boldtheta}_S}) \right] + 2\mathbb{E}_{\cPN}\left[ \hat{R}_{S}(\boldf_{\boldtheta^*})- R(\boldf_{\boldtheta^*})\right] \nonumber \\
		~ &=& \mathbb{E}_{\cPN}\left[ R(\boldf_{\boldtheta^*})-2\hat{R}_{S}(\boldf_{\hat{\boldtheta}_S})+R(\boldf_{\hat{\boldtheta}_S}) \right],
	\end{eqnarray}
	where the inequality holds since $\hat{R}_{S}(\boldf_{\hat{\boldtheta}_S})\leq \hat{R}_{S}(\boldf_{\boldtheta^*})$ and the last equality holds since $\boldf_{\boldtheta^*}$ does not depend on $S$. 
	
	
	We further reformulate the expectation in~Eq.~(\ref{eqn:sta_err}) by introducing the ghost sample. Namely, let $S^{\prime}=\left\{\left(\bfx_i^{\prime}, y_i^{\prime}\right)\right\}_{i=1}^N$ be an i.i.d.~copy of $S=\left\{\left(\bfx_i, y_i\right)\right\}_{i=1}^N$, and let the density of the sample pairs as $\cP^{\prime}$. Denote $z:=\left(\bfx, y\right)$ and $z^{\prime}:=\left(\bfx^{\prime}, y^{\prime}\right)$. We define the following two random variables depending on $z$:
	\begin{eqnarray}\label{eqn:def_g}
		g(\boldf_{\boldtheta}, z) = \ell(\boldf_{\boldtheta}, z) - \ell(\boldf_{\boldtheta^{*}}, z), 
	\end{eqnarray}
	and
	\begin{eqnarray}\label{eqn:def_G}
		G(\boldf_{\boldtheta}, z) = \mathbb{E}_{\cP^{\prime}}\left[ g(\boldf_{\boldtheta}, z^{\prime})-2g(\boldf_{\boldtheta}, z) \right]=\mathbb{E}_{\cP^{\prime}}\left[ g(\boldf_{\boldtheta}, z^{\prime}) \right]-2g(\boldf_{\boldtheta}, z).
	\end{eqnarray}
	We establish the following lemma.
	\begin{lemma}\label{lem:reformulate}
		\begin{equation}\label{eq: reformulate}
			\mathbb{E}_{\cPN}\left[ R(\boldf_{\boldtheta^{*}})-2\hat{R}_{S}(\boldf_{\hat{\boldtheta}_S})+R(\boldf_{\hat{\boldtheta}_S}) \right] = \mathbb{E}_{\cPN}\left[ \frac{1}{N}\sum_{i=1}^{N}G(\boldf_{\hat{\boldtheta}_S}, z_i) \right].
		\end{equation}
	\end{lemma} 
	\begin{proof}
		Note that the distribution over $\boldf_{\hat{\boldtheta}_S}$ is induced from the random samples $S\sim \cP^{\otimes N}$. Since the density $\cP'$ of the ghost samples is independent with $\cP$, it is also independent with the distribution over $\boldf_{\hat{\boldtheta}_S}$. Thus, $G(\boldf_{\boldtheta}, z) = \mathbb{E}_{\cP^{\prime}}\left[ g(\boldf_{\boldtheta}, z^{\prime})\right]-2g(\boldf_{\boldtheta}, z)$ and we have
		\begin{eqnarray}
			\mathbb{E}_{\cPN}\left[ \frac{1}{N}\sum_{i=1}^{N}G(\boldf_{\hat{\boldtheta}_S}, z_i) \right] &=& \mathbb{E}_{\cPN}\left[ \frac{1}{N}\sum_{i=1}^{N}\left( \mathbb{E}_{\cP^{\prime\otimes N}}\left[ g(\boldf_{\hat{\boldtheta}_S}, z_i^{\prime}) \right]-2g(\boldf_{\hat{\boldtheta}_S}, z_i) \right) \right] \nonumber \\
			~ &=& \mathbb{E}_{\cPN}\left[ \mathbb{E}_{\cP^{\prime\otimes N}}\left[ g(\boldf_{\hat{\boldtheta}_S}, z^{\prime}) \right]-\frac{2}{N}\sum_{i=1}^{N}g(\boldf_{\hat{\boldtheta}_S}, z_i) \right] \nonumber \\
			~ &=& \mathbb{E}_{\cPN}\left[ \mathbb{E}_{\cP^{\prime\otimes N}}\left[ \ell(\boldf_{\hat{\boldtheta}_S}, z^{\prime}) - \ell(\boldf_{{\boldtheta}^{*}}, z^{\prime}) \right]-\frac{2}{N}\sum_{i=1}^{N}\left( \ell(\boldf_{\hat{\boldtheta}_S}, z_{i}) - \ell(\boldf_{{\boldtheta}^{*}}, z_{i}) \right) \right] \nonumber \\
			~ &=& \mathbb{E}_{\cPN}\left[ R(\boldf_{\hat{\boldtheta}_S})-R(\boldf_{{\boldtheta}^{*}})-2\hat{R}_{S}(\boldf_{\hat{\boldtheta}_S})+2\hat{R}_{S}(\boldf_{{\boldtheta}^{*}}) \right] \nonumber \\
			~ &=& \mathbb{E}_{\cPN}\left[ R(\boldf_{\hat{\boldtheta}_S})-R(\boldf_{{\boldtheta}^{*}})-2\hat{R}_{S}(\boldf_{\hat{\boldtheta}_S})+2{R}(\boldf_{{\boldtheta}^{*}}) \right] \nonumber \\
			~ &=& \mathbb{E}_{\cPN}\left[ R(\boldf_{\boldtheta^{*}})-2\hat{R}_{S}(\boldf_{\hat{\boldtheta}_S})+R(\boldf_{\hat{\boldtheta}_S}) \right], \nonumber
		\end{eqnarray}
		where the second last equality holds since $\mathbb{E}_{\cP^{\otimes N}}\left[ \hat{R}_{S}(\boldf_{\boldtheta^*}) \right] = {R}(\boldf_{\boldtheta^*}) = \mathbb{E}_{\cP^{\otimes N}}\left[ {R}(\boldf_{\boldtheta^*}) \right]$. 
	\end{proof}
	Lemma~\ref{lem:reformulate} helps to rewrite Eq.~(\ref{eqn:sta_err}) as the expectation of $N$ random variables; in other words, we can estimate the expected prediction error by integrating the tail probability 
	\[
	\mathbb{P}\left(\frac{1}{N}\sum_{i=1}^{N}G(\boldf_{\hat{\boldtheta}_S}, z_i)>t \right).
	\]
	However, directly estimating $\mathbb{P}\left(\frac{1}{N}\sum_{i=1}^{N}G(\boldf_{\hat{\boldtheta}_S}, z_i)>t \right)$ is challenging since we do not have access to the distribution of $G(\boldf_{\hat{\boldtheta}_S}, z_i)$. Thus, we can upper bound this quantity by upper bounding 
	\[
	\mathbb{P}\left(\sup \limits_{\boldf_{\boldtheta}\in\cH }\frac{1}{N}\sum_{i=1}^{N}G(\boldf_{\boldtheta}, z_i)>t \right).
	\] 
	This can be done by the following lemma, which utilizes an $\varepsilon$-covering net of $\cH$.
	
	\begin{lemma}\label{lem:cover}
		Let $\mathcal{B}$ be a constant which upper bounds $|y|$, $|\boldf_{{\boldtheta}}(\bfx)|$ and $\|\cO\|$ for any $\bfx\in \cX$, $y\in\cY$ and $\boldf_{{\boldtheta}}\in\cH$. For any $t\geq 0$, we have 
		$$\mathbb{P}\left(\sup \limits_{\boldf_{\boldtheta}\in\cH }\frac{1}{N}\sum_{i=1}^{N}G(\boldf_{\boldtheta}, z_i)>t \right)\leq \mathbb{P}\left(\max \limits_{\boldf_{\boldtheta_{\varepsilon}}\in\mathcal{N}(\mathcal{H}, \varepsilon, |\cdot|)} \frac{1}{N}\sum_{i=1}^{N}G(\boldf_{\boldtheta_{\varepsilon}}, z_i)>t-12\mathcal{B}\varepsilon\right),$$ 
		where $\mathcal{N}(\mathcal{H}, \varepsilon, |\cdot|)$ is an $\varepsilon$-covering net of $\mathcal{H}$ with respect to the absolute norm $|\cdot|$.
	\end{lemma}
	\begin{proof}
		Let $\mathcal{N}(\mathcal{H}, \varepsilon, |\cdot|)$ be an $\varepsilon$-covering net of $\mathcal{H}$. For any $\boldf_{\boldtheta}\in\cH$, there exists $\boldf_{\boldtheta_{\varepsilon}}\in \mathcal{N}(\mathcal{H}, \varepsilon, |\cdot|)$ satisfying $| \boldf_{\boldtheta}-\boldf_{\boldtheta_{\varepsilon}} |\leq \varepsilon$.
		Thus, for any $z=(\bfx,y)$
		\begin{eqnarray}\label{eqn:g_theta}
			\left|g(\boldf_{\boldtheta}, z)-g(\boldf_{\boldtheta_{\varepsilon}}, z) \right|=\left|\ell(\boldf_{\boldtheta}, z)-\ell(\boldf_{\boldtheta_{\varepsilon}}, z) \right| = \left|(\boldf_{\boldtheta}(\bfx)-y)^2-(\boldf_{\boldtheta_{\varepsilon}}(\bfx)-y)^2 \right| 
			&\leq& 4\mathcal{B} \left| \boldf_{\boldtheta}(\bfx)-\boldf_{\boldtheta_{\varepsilon}}(\bfx) \right| \nonumber \\ &\leq& 4\mathcal{B} \varepsilon. \nonumber
		\end{eqnarray}
		Moreover, we have
		\begin{eqnarray}
			G(\boldf_{\boldtheta}, z)-G(\boldf_{\boldtheta_{\varepsilon}}, z) &=& \mathbb{E}_{\cP^{\prime}}\left[  g(\boldf_{\boldtheta}, z^{\prime})-g(\boldf_{\boldtheta_{\varepsilon}}, z^{\prime})\right]-2\left( g(\boldf_{\boldtheta}, z)-g(\boldf_{\boldtheta_{\varepsilon}}, z) \right)  \leq 12\mathcal{B} \varepsilon. \nonumber
		\end{eqnarray}
		Since $\mathcal{N}(\mathcal{H}, \varepsilon, |\cdot|)$ is a finite set, we have $$\sup \limits_{\boldf_{\boldtheta}\in\cH}\frac{1}{N}\sum_{i=1}^{N}G(\boldf_{\boldtheta}, z_i) \leq \max \limits_{\boldf_{\boldtheta_{\varepsilon}}\in\mathcal{N}(\mathcal{H}, \varepsilon, |\cdot|)} \frac{1}{N}\sum_{i=1}^{N}G(\boldf_{\boldtheta_{\varepsilon}}, z_i)+12\mathcal{B} \varepsilon, $$ 
		which leads to 
		\begin{eqnarray}
			\mathbb{P}\left(\sup \limits_{\boldf_{\boldtheta}\in\cH }\frac{1}{N}\sum_{i=1}^{N}G(\boldf_{\boldtheta}, z_i)>t \right)\leq \mathbb{P}\left(\max \limits_{\boldf_{\boldtheta_{\varepsilon}}\in\mathcal{N}(\mathcal{H}, \varepsilon, |\cdot|)} \frac{1}{N}\sum_{i=1}^{N}G(\boldf_{\boldtheta_{\varepsilon}}, z_i)>t-12\mathcal{B}\varepsilon\right).\nonumber
		\end{eqnarray}
	\end{proof}
	Meanwhile, we can upper bound the tail probability using Bernstein inequality (cf.~\Cref{thm:Bernstein_ineq}). 
	\begin{lemma}\label{lem:Bernstein}
		Let $\mathcal{B}$ be a constant which upper bounds $|y|$, $|\boldf_{{\boldtheta}}(\bfx)|$ and $\|\cO\|$ for any $\bfx\in \cX$, $y\in\cY$ and $\boldf_{{\boldtheta}}\in\cH$. For any $t\geq 0$ and $\boldf_{\boldtheta}\in\cH$, we have 
		$$\mathbb{P}\left( \frac{1}{N}\sum_{i=1}^{N}G(\boldf_{\boldtheta}, z_i)\geq t \right)\leq \exp \left( -\frac{Nt}{320\mathcal{B}^2} \right),$$
		where $\boldf_{\boldtheta_{\varepsilon}}\in\mathcal{N}(\mathcal{H}, \varepsilon, |\cdot|)$.
	\end{lemma}
	\begin{proof}
		We first bound the bias and variance of the random variable $g(\boldf_{\boldtheta}, z)$. Note that $|\ell(\boldf_{\boldtheta}, z)|=(\boldf_{\boldtheta}(\bfx)-y)^2 \leq 4\mathcal{B}^2$. We have $|g(\boldf_{\boldtheta}, z)|=|\ell(\boldf_{\boldtheta}, z)-\ell(\boldf_{\boldtheta^*}, z)|\leq 8\mathcal{B}^2$, and
		\begin{eqnarray}\label{eqn:Bernstein_ineq_0}
			b_1:=|g(\boldf_{\boldtheta}, z)-\mathbb{E}_\cP[g(\boldf_{\boldtheta}, z)]|\leq 16\mathcal{B}^2.
		\end{eqnarray}
		
		Denote the variance of $g(\boldf_{\boldtheta}, z)$ be $\sigma_1^2$. We have 
		\[\sigma_1^2\leq \mathbb{E}_{\cP}\left[g(\boldf_{\boldtheta}, z)^2\right]\leq (4\mathcal{B})^2\mathbb{E}_{\cP}\left[\left(\boldf_{\boldtheta}(\bfx)-\boldf_{\boldtheta^*}(\bfx)\right)^2\right]\leq 32\mathcal{B}^2\mathbb{E}_{\cP}\left[\ell(\boldf_{\boldtheta}, z)-\ell(\boldf_{\boldtheta^*}, z) \right] = 32\mathcal{B}^2 \mathbb{E}_\cP[g(\boldf_{\boldtheta}, z)], \] 
		where the third inequality utilizes the fact that (1) \(\ell(\boldf_{\boldtheta}, z)\) is strongly convex with coefficient \(1/2 \) and (2) \(\boldf_{\boldtheta^*}\) is a global minimum of \(\ell(\boldf_{\boldtheta}, z)\). Thus, we have that 
		\begin{eqnarray}\label{eqn:Bernstein_ineq_1}
			\mathbb{E}_\cP[g(\boldf_{\boldtheta}, z)] \geq \frac{\sigma_1^2}{32\mathcal{B}^2}. 
		\end{eqnarray}
		Now for any $t\geq 0$, we have that
		\begin{eqnarray}
			~ &~&\mathbb{P}\left( \frac{1}{N}\sum_{i=1}^{N}G(\boldf_{\boldtheta}, z_i)\geq t \right) \nonumber \\
			&=& \mathbb{P}\left( \mathbb{E}_{\cP^{\prime\otimes N}}\left[ \frac{1}{N}\sum_{i=1}^{N} g(\boldf_{\boldtheta}, z_i^{\prime}) \right]-\frac{2}{N}\sum_{i=1}^{N} g(\boldf_{\boldtheta}, z_i) \geq t \right) \nonumber \\
			&=& \mathbb{P}\left( \mathbb{E}_{\cPN}\left[ \frac{1}{N}\sum_{i=1}^{N} g(\boldf_{\boldtheta}, z_i) \right]-\frac{1}{N}\sum_{i=1}^{N} g(\boldf_{\boldtheta}, z_i) \geq \frac{t}{2}+ \mathbb{E}_{\cPN}\left[ \frac{1}{2N}\sum_{i=1}^{N} g(\boldf_{\boldtheta}, z_i) \right] \right) \nonumber \\
			&\leq& \mathbb{P}\left( \mathbb{E}_{\cPN}\left[ \frac{1}{N}\sum_{i=1}^{N} g(\boldf_{\boldtheta}, z_i) \right]-\frac{1}{N}\sum_{i=1}^{N} g(\boldf_{\boldtheta}, z_i) \geq \frac{t}{2}+\frac{\sigma_1^2}{64\mathcal{B}^2} \right),  \nonumber 
		\end{eqnarray}
		where the first equality uses the fact that $\mathbb{E}_{\cPN}\left[ \frac{1}{N}\sum_{i=1}^{N} g(\boldf_{\boldtheta}, z_i)\right]=\mathbb{E}_{\cP^{\prime\otimes N}}\left[ \frac{1}{N}\sum_{i=1}^{N} g(\boldf_{\boldtheta}, z_i^{\prime}) \right]$, since an arbitrary $\boldtheta$ is independent with both $\cP$ and $\cP'$. Set $u=t/2+\sigma_1^2/{(64\mathcal{B}^2)}$, we have $\sigma_1^2\leq 64\mathcal{B}^2 u$ and $u\geq t/2$. Utilizing the Bernstein inequality in~\Cref{thm:Bernstein_ineq}, we obtain
		\begin{eqnarray}
			\mathbb{P}\left( \frac{1}{N}\sum_{i=1}^{N}G(\boldf_{\boldtheta}, z_i)\geq t \right) &\leq& \mathbb{P}\left( \mathbb{E}_{\cPN}\left[ \frac{1}{N}\sum_{i=1}^{N} g(\boldf_{\boldtheta}, z_i) \right]-\frac{1}{N}\sum_{i=1}^{N} g(\boldf_{\boldtheta}, z_i) \geq u \right)  \nonumber \\
			&\leq& \exp \left( -\frac{Nu^2}{2(\sigma_1^2+b_1u)} \right)  \nonumber \\
			&\leq& \exp \left( -\frac{Nu^2}{2(64\mathcal{B}^2u+16\mathcal{B}^2u)} \right)  \nonumber \\
			&\leq& \exp \left( -\frac{Nt}{320\mathcal{B}^2} \right).  \nonumber 
		\end{eqnarray}
	\end{proof}
	Now we are ready to prove our first main theorem:
	\begin{theorem}\label{thm1_upper_bound}
		Consider a linear QML model $(V(\bfx),U(\boldtheta),\cO)$, where $V(\bfx)$ is an arbitrary feature map acting on $n$-qubits, $U_{\boldtheta}$ is a parametrized quantum circuit consisting of $T$ parameterized quantum gates acting on at most $2$-qubits and an arbitrary number of non-trainable quantum gates, and $\cO$ is a measurement observable with bounded operator norm. Let $\mathcal{B}$ be a constant which upper bounds $|y|$, $|\boldf_{{\boldtheta}}(\bfx)|$ and $\|\cO\|$ for any $\bfx\in \cX$, $y\in\cY$, and $\boldf_{{\boldtheta}}\in\cH$. We have
		\begin{eqnarray}\label{sm_eqn_thm1}
			\mathbb{E}_{\cP^{\otimes N}}\left[\cE_P(\boldf_{\hat{\boldtheta}_S}) \right]\leq\frac{320\mathcal{B}^2}{N} \left( 16T\left(\log{(\frac{21N}{1280T})}+1 \right)+1 \right)=\tilde{O}(T/N). 
		\end{eqnarray}
		Ignoring the logarithmic factors in Eq.~(\ref{sm_eqn_thm1}), we can simplify it as $\mathbb{E}\left[\cE_P(\boldf_{\hat{\boldtheta}_S}) \right]\leq	 C\cdot T/N$ for the constants $C>0$. Since $\cE_P(\boldf_{\hat{\boldtheta}_S})\in[0, 4\mathcal{B}^2]$, from Hoeffding's inequality (cf.~\Cref{thm:Hoeffding_ineq_1}) we have that for any $t>0$,
		\begin{eqnarray}\label{sm_eqn_thm1_P}
			\bP\left(\cE_P(\boldf_{\hat{\boldtheta}_S})\geq C\cdot T/N+t \right)\leq\exp{-\frac{2t^2}{16\mathcal{B}^4}}. 
		\end{eqnarray}
		Thus we state that, for almost all training data set $S$ of size $N$, we have $\cE_P(\boldf_{\hat{\boldtheta}_S})=\tilde{O}(T/N)$.
	\end{theorem}
	\begin{proof}
		Note that 
		\begin{eqnarray}
			\mathbb{E}_{\cPN}\left[ \cE_P(\boldf_{\hat{\boldtheta}_S}) \right] &\leq& \mathbb{E}_{\cPN}\left[ R(\boldf_{\boldtheta^*})-2\hat{R}_{S}(\boldf_{\hat{\boldtheta}_S})+R(\boldf_{\hat{\boldtheta}_S}) \right] \text{(using Eq.~(\ref{eqn:sta_err}))} \nonumber \\
			~ &=& \mathbb{E}_{\cPN}\left[ \frac{1}{N}\sum_{i=1}^{N}G(\boldf_{\hat{\boldtheta}_S}, z_i) \right] \text{ (by Lemma~\ref{lem:reformulate})} \nonumber \\
			~ &\leq& \int_{0}^{+\infty} \mathbb{P}\left(\frac{1}{N}\sum_{i=1}^{N}G(\boldf_{\hat{\boldtheta}_S}, z_i)>t \right)dt  \nonumber \\ 
			~ &\leq&  \int_{0}^{+\infty} \mathbb{P}\left(\sup \limits_{\boldf_{\boldtheta}\in\cH}\frac{1}{N}\sum_{i=1}^{N}G(\boldf_{{\boldtheta}}, z_i)>t \right)dt,  \nonumber \\ 
		\end{eqnarray}
		where the second inequality uses the fact that if $\mathbb{E}\left[ \left|W\right| \right]<+\infty$, we have $\mathbb{E}\left[ W \right]=\int_{0}^{+\infty}\mathbb{P}\left(W>w\right)dw-\int_{-\infty}^{0}\mathbb{P}\left(W\leq w\right)dw \leq \int_{0}^{+\infty}\mathbb{P}\left(W>w\right)dw$. 
		Let $c={(320\mathcal{B}^2\log{|\mathcal{N}(\mathcal{H}, \varepsilon, |\cdot|)|})}/{N}+12\mathcal{B}\varepsilon$, where $\varepsilon=1280T\mathcal{B}/(3N)$. We further have
		\begin{eqnarray}
			\mathbb{E}_{\cPN}\left[ \cE_P(\boldf_{\hat{\boldtheta}_S}) \right] &\leq&  \int_{0}^{+\infty} \mathbb{P}\left(\sup \limits_{\boldf_{\boldtheta}\in\cH}\frac{1}{N}\sum_{i=1}^{N}G(\boldf_{{\boldtheta}}, z_i)>t \right)dt,  \nonumber \\ 
			~ &\leq& \int_{0}^{c} \mathbb{P}\left(\max \limits_{\boldf_{\boldtheta_{\varepsilon}}\in\mathcal{N}(\mathcal{H}, \varepsilon, |\cdot|)} \frac{1}{N}\sum_{i=1}^{N}G(\boldf_{\boldtheta_{\varepsilon}}, z_i)>t-12\mathcal{B}\varepsilon\right)dt  \nonumber \\ 
			~ &+&  \int_{c}^{+\infty} \mathbb{P}\left(\max \limits_{\boldf_{\boldtheta_{\varepsilon}}\in\mathcal{N}(\mathcal{H}, \varepsilon, |\cdot|)} \frac{1}{N}\sum_{i=1}^{N}G(\boldf_{\boldtheta_{\varepsilon}}, z_i)>t-12\mathcal{B}\varepsilon\right)dt \text{ (by Lemma~\ref{lem:cover})}  \nonumber \\ 
			~ &\leq& c+ \int_{c}^{+\infty} \left|\mathcal{N}(\mathcal{H}, \varepsilon, |\cdot|) \right| \mathbb{P}\left(\frac{1}{N}\sum_{i=1}^{N}G(\boldf_{\boldtheta_{\varepsilon}}, z_i)>t-12\mathcal{B}\varepsilon \right)dt \nonumber \\ 
			~ &\leq& c+ \int_{c}^{+\infty} \left|\mathcal{N}(\mathcal{H}, \varepsilon, |\cdot|) \right| \exp \left( -\frac{N(t-12\mathcal{B}\varepsilon)}{320\mathcal{B}^2} \right)dt \text{ (by Lemma~\ref{lem:Bernstein})} \nonumber \\ 
			~ &\leq& \frac{320\mathcal{B}^2}{N} \left( \log{(\left|\mathcal{N}(\mathcal{H}, \varepsilon, |\cdot|) \right|)} + 3N\varepsilon/(80\mathcal{B}) +1 \right) \nonumber \\ 
			~ &\leq& \frac{320\mathcal{B}^2}{N} \left( 16T\left(\log{(\frac{21N}{1280T})}+1 \right)+1 \right) \text{ (by Lemma~\ref{corollary:covering-number-abs-norm})}. \nonumber 
		\end{eqnarray}		
	\end{proof}

	\subsection{Extensions to data re-uploading QML models}
\begin{figure*}[ht!]
		\centering
		\includegraphics[width=0.9\textwidth]{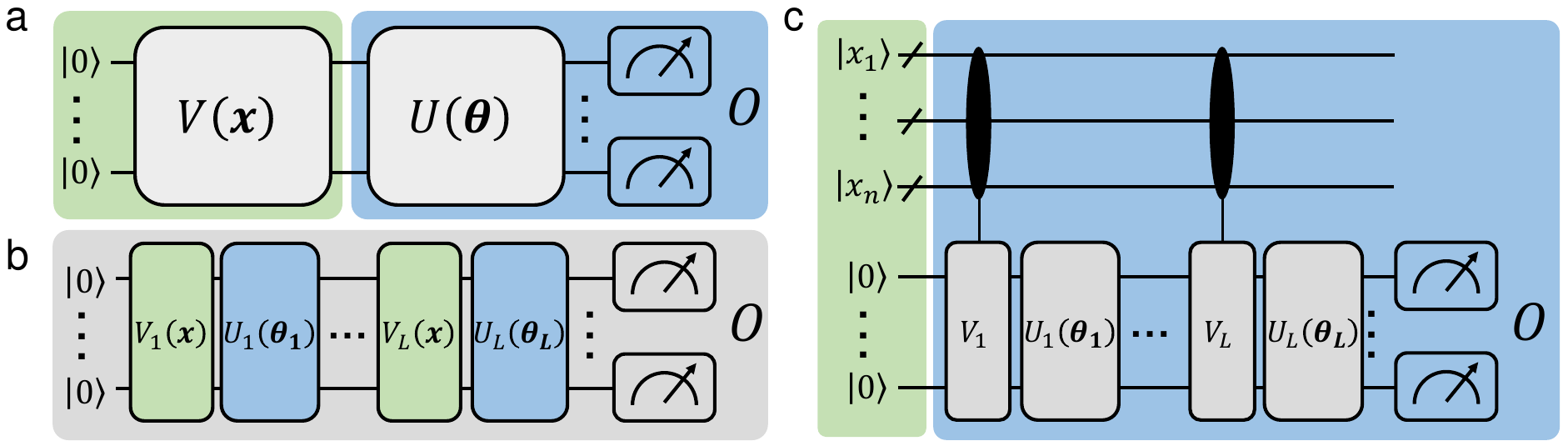}
		\caption{\small{\textbf{Overview the QML models.} (a) Linear quantum models. (b) Data re-uploading models. (c) An equivalent representation of the data re-uploading models. Every quantum register corresponds to $q$ quantum qubits. The green background squares depict the data encoding unitaries, while the blue background squares represent the unitaries subject to optimization. See also \cite{jerbiQuantumMachineLearning2023a} for reference.
		}}
		\label{fig:QML-models}
	\end{figure*}

	\paragraph{Data re-uploading QML models.} Before the introduction of data re-uploading QML models, recall that the linear QML model is defined by a data-encoding circuit \(V(\bfx)\), a trainable parameterized quantum circuit \(U(\boldtheta)\) and a prefixed observable \(\cO\). The output of the linear QML model is denoted by \(\Tr[U_{\boldtheta}^{\dagger}\cO U_{\boldtheta}\rho(\bfx)]\), where $\rho(\bfx)=V(\bfx)\ket{0}\!\bra{0}V^{\dagger}(\bfx)$. A visualization of the linear QML model can be seen in FIG.~\ref{fig:QML-models}(a).
	
	Different from the linear QML model, the data re-uploading QML model interweaves the data-encoding circuits \(V_1(\bfx), V_2(\bfx), \dots, V_L(\bfx) \) and trainable circuits \(U_1(\boldtheta), U_2(\boldtheta), \dots, U_L(\boldtheta)\) to construct the circuit model.  We define the restrictive data re-uploading QML model as:
	\begin{itemize}
		\item The data point $\bfx:=[x_1, x_2, \dots, x_n]^T\in[0, 1]^n$. Each component \(x_i\) has an exact \(q\)-bit representation. In other words, for \(\forall i\in[n]\), there exists a bit string \(b_1b_2\cdots b_{q} \in\{0, 1\}^q\) such that \(x_i=b_12^{-1}+b_22^{-2}+\cdots +b_q2^{-q}\).
		\item The \(n\)-qubit data-encoding circuits \(V_1(\bfx), V_2(\bfx), \dots, V_L(\bfx) \). Without loss of generality, each circuit \(V_i(\bfx)\) is composed of \(\Rz(x_1)\otimes \Rz(x_2)\otimes\cdots\otimes \Rz(x_n)\), where \(\Rz(x)=\left(\begin{smallmatrix}e^{-ix/2}&0\\0&e^{ix/2}\end{smallmatrix}\right)\).
		\item The \(n\)-qubit trainable circuits \(U_1(\boldtheta_1), U_2(\boldtheta_2), \dots, U_L(\boldtheta_L)\) with \(\boldtheta=[\boldtheta_1, \dots, \boldtheta_L]\in(0, 2\pi]^T\).
		\item A prefixed observable \(\cO\) with bounded operator norm.
	\end{itemize}
	The original definition of data re-uploading QML model proposed by~\citet{PerezSalinas2020datareuploading} excludes the first restriction. We make this restriction to facilitate the calculation of the covering entropy of the space of data re-uploading QML models.
	The output of a data re-uploading QML model is denoted
	\[
	\tilde{\boldf}_{\boldtheta}(\bfx)=\Tr[U_{\boldtheta}(\bfx)^{\dagger}\cO U_{\boldtheta}(\bfx)\rho(\bfx)],
	\]
	where \(U_{\boldtheta}(\bfx)=U_L(\boldtheta_L)V_L(\bfx)\cdots U_2(\boldtheta_2)V_2(\bfx)U_1(\boldtheta_1) \) and \(\rho(\bfx)=V_1(\bfx)\ket{0}\!\bra{0}V_1^{\dagger}(\bfx)\).
	A visualization of the data re-uploading QML model can be seen in FIG.~\ref{fig:QML-models}(b).
	
	\paragraph{The covering number of data re-uploading QML models.}
	To measure the covering number of the data re-uploading QML model, we shall transform it into a linear QML model, albeit within an expanded quantum system. Subsequently, we employ~\Cref{thm:covering_number_circuit} to establish the covering number upper bound of the linear QML model.
	
\begin{figure*}[ht!]
		\centering
		\includegraphics[width=0.5\textwidth]{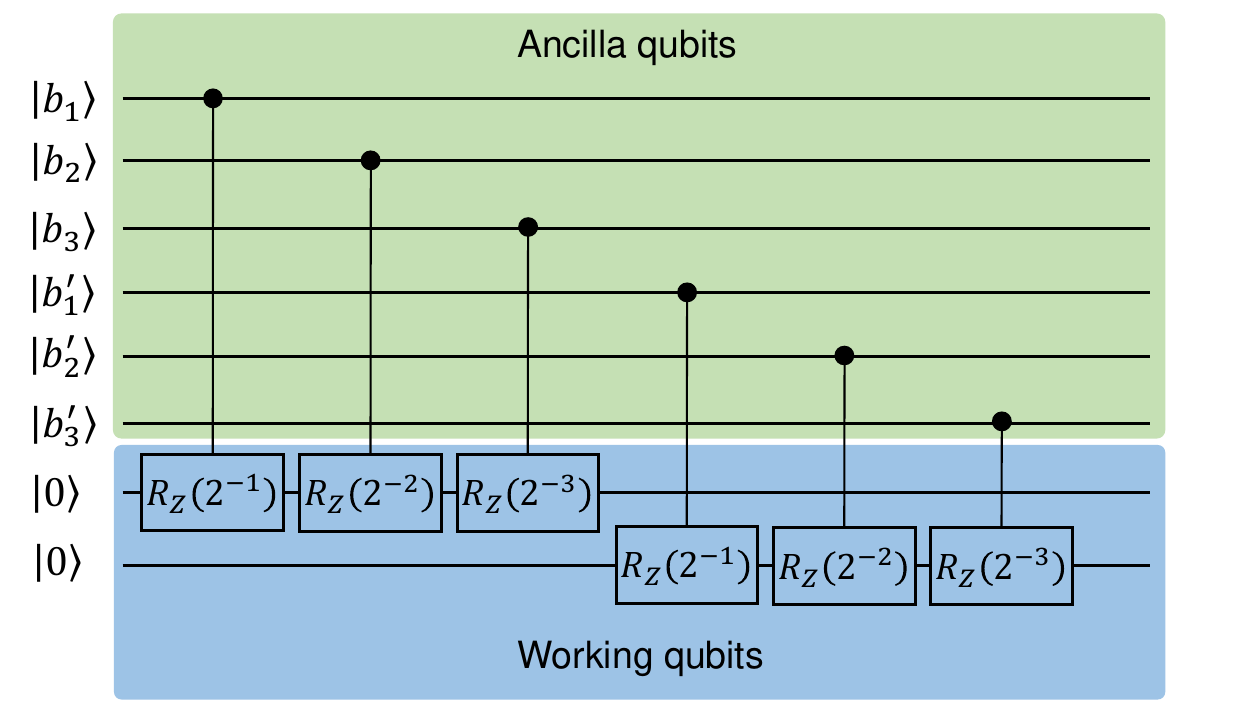}
		\caption{\small{\textbf{A template circuit of the \(\text{C-}V_i\) operator.} We consider a special example of \(n=2\) qubits QML model and \(q=3\). Let \(\bfx=(x_1, x_2)\), \(\ket{x_1}=\ket{b_1b_2b_3}\) and \(\ket{x_2}=\ket{b_1'b_2'b_3'}\). The \(\text{C-}V_i\) circuit is composed of the product of \(n=2~\cRz\) operators.  
		}}
		\label{fig:data_re_uploading}
	\end{figure*}

	To elaborate, consider input data $\bfx=[x_1, x_2, \dots, x_n]^T$. We can encode \(x_i\) as the bit string \(\ket{x_i}=\ket{b_1b_2\cdots b_q}\in\{0, 1\}^q\). Denote these data-encoding qubits as ancilla qubits. The data \(\bfx\) can be encoded into \(dq\) ancilla qubits. To derive the effect of the data-encoding unitary $\Rz(x_i)$ on working qubits. We can employ \(q\) \(\Rz\) gates with fixed rotation angles, i.e., \(\Rz(2^{-1}), \Rz(2^{-2}), \dots, \Rz(2^{-q})\), controlled by \(q\) ancilla qubits, i.e., \(\ket{b_1b_2\cdots b_q}\). Since 
	\[ \Rz(b_1 2^{-1}) \Rz(b_2 2^{-2}) \cdots \Rz(b_{q} 2^{-q}) =\Rz(b_1 2^{-1} + b_2 2^{-2} + \cdots + b_{q} 2^{-q}) = \Rz(x_i), \]
	we can realize the effect of a data-encoding gate \(\Rz(x_i)\) by the product of Control-\(\Rz\) (C-$\Rz$) gates. The complete \((q+1)\)-qubit circuit is denoted as \({\cRz}=\prod_{i=1}^{q}\text{C}_i\text{-}\Rz(2^{-i})\), where \(\text{C}_i\) represents control qubit \(b_i\). Different from \(\Rz(x_i)\) dependent on the sample data \(x_i\), \(\cRz\) is a fixed quantum circuit independent on \(x_i\). In this way, we can use an \(n(q+1)\)-qubit circuit \(\text{C-}V_i\) to represent the \(V_i(\bfx)\) operator \(\forall i\in[n]\), with \(nq\) ancilla qubits. An illustration of \(\cRz\) operator and \(\text{C-}V_i\) operator is shown in FIG.~\ref{fig:data_re_uploading}.
	
	This insight provides the linear QML model representation \((V(\bfx), U(\boldtheta), \cO)\) of data re-uploading QML models as below:
	\begin{itemize}
		\item The data point $\bfx:=[x_1, x_2, \dots, x_n]^{\top}\in[0, 1]^n$. Each component \(x_i\) has an exact \(q\)-bit representation.
		\item The \(n(q+1)\)-qubit data-encoding operator \(V(\bfx)\) has effect \(V(\bfx)\ket{0^{\otimes n(q+1)}}=\ket{0^{\otimes n}}\ket{x_1}\ket{x_2}\cdots\ket{x_n}\), whose density matrix representation is denoted as \(\tilde\rho(\bfx)=V(\bfx)\ket{0^{\otimes n(q+1)}}\!\bra{0^{\otimes n(q+1)}}V(\bfx)^{\dagger}\).
		\item The \(n(q+1)\)-qubit trainable circuits \(U(\boldtheta)=\text{C-}V_1(\bI\otimes U_1(\boldtheta_1))\cdots \text{C-}V_L(\bI\otimes U_L(\boldtheta_L)) \).
		\item A prefixed observable \(\cO\) with bounded operator norm.
	\end{itemize}
	The output of this linear QML model is exactly the output of the data re-uploading QML model, which is denoted as \(\tilde\boldf_{\boldtheta}(\bfx)=\Tr[\cO\Tr_1[U(\boldtheta)\tilde{\rho}(\bfx)U(\boldtheta)^{\dagger}]]\), where \(\Tr_1\) denotes partial trace the effect of ancilla qubits. A visualization of this linear QML model can be seen in FIG.~\ref{fig:QML-models}(c).
	
	Using the established covering entropy upper bound of linear QML models (cf.~\Cref{thm:covering_number_circuit}), we have that for any fixed data point \(\bfx\), the hypothesis space \(\{U(\boldtheta)\tilde{\rho}(\bfx)U(\boldtheta)^{\dagger}:~\boldtheta\in(0, 2\pi]^T\}\) has \(\varepsilon\)-covering entropy \(O(T\log(1/\varepsilon))\) with respective to the operator norm \(\|\cdot\|\). We denote its covering set as  \(\mathcal{N}=\{U(\boldtheta_{\varepsilon})\tilde{\rho}(\bfx)U(\boldtheta_{\varepsilon})^{\dagger}\).
    Let 
	\[
	\tilde{\mathcal{N}}=\left\{\tilde\boldf_{\boldtheta_\varepsilon}(\bfx)=\Tr[\cO\Tr_1[U(\boldtheta_{\varepsilon})\tilde{\rho}(\bfx)U(\boldtheta_{\varepsilon})^{\dagger}]]:~U(\boldtheta_{\varepsilon})\tilde{\rho}(\bfx)U(\boldtheta_{\varepsilon})^{\dagger}\in\mathcal{N}\right\}.
	\]
	It holds that 
	\begin{eqnarray}
	|\tilde\boldf_{{\boldtheta}}(\bfx)-\tilde\boldf_{\boldtheta_{\varepsilon}}(\bfx)| &=& \bigl|\Tr[\cO\Tr_1[U(\boldtheta)\tilde{\rho}(\bfx)U(\boldtheta)^{\dagger} - U(\boldtheta_{\varepsilon})\tilde{\rho}(\bfx)U(\boldtheta_{\varepsilon})^{\dagger}]]\bigr| \nonumber \\
	~ &\leq& \|\cO\| \bigl|\Tr[\Tr_1[U(\boldtheta)\tilde{\rho}(\bfx)U(\boldtheta)^{\dagger} - U(\boldtheta_{\varepsilon})\tilde{\rho}(\bfx)U(\boldtheta_{\varepsilon})^{\dagger}]]\bigr| \nonumber \\
	~ &\leq& \|\cO\|2^{n(q+1)} \|U(\boldtheta)\tilde{\rho}(\bfx)U(\boldtheta)^{\dagger} - U(\boldtheta_{\varepsilon})\tilde{\rho}(\bfx)U(\boldtheta_{\varepsilon})^{\dagger}\|,\nonumber
	\end{eqnarray}
	which means that \(\tilde{\mathcal{N}}\) forms a \(\|\cO\|2^{n(q+1)}\varepsilon\)-covering set of \(\mathcal{\tilde{H}}=\{\tilde\boldf_{\boldtheta}(\bfx):~\boldtheta\in(0, 2\pi]^T\}\). This implies that the \(\varepsilon\)-covering entropy of \(\mathcal{\tilde{H}}\) scales as \(O\left(T(\log(1/\varepsilon)+nq)\right)\). Summarizing the above, we have the following:
	
	\begin{lemma}\label{lemma_covering_re_uploading}
		(Covering entropy upper bound for \(\mathcal{\tilde{H}}\) with respective to operator norm) Suppose the data point $\bfx:=[x_1, x_2, \dots, x_n]^T\in[0, 1]^n$ and each component \(x_i\) has an exact \(q\)-bit representation. Let \(V_1(\bfx), V_2(\bfx), \dots, V_L(\bfx)\) denote the data encoding circuits, with each component \(V_i(\bfx)=\otimes_{i=1}^n\Rz(x_i)\). Let \(U_1(\boldtheta_1), U_2(\boldtheta_2), \dots, U_L(\boldtheta_L)\) with \(\boldtheta=[\boldtheta_1, \dots, \boldtheta_L]\in(0, 2\pi]^T\) denote the trainable quantum circuit. Let \(\cO\) be a prefixed observable with a bounded operator norm. The data re-uploading QML model is defined by \((V_1(\bfx),\dots, V_L(\bfx), U_1(\boldtheta_1),\dots, U_L(\boldtheta_L), \cO)\). Then for \(\forall\varepsilon\in[0, 0.1)\), the \(\varepsilon\)-covering entropy of \(\mathcal{\tilde{H}}\) with respective to \(|\cdot|\) satisfies:
		\[
		\log(\mathcal{N}(\mathcal{\tilde{H}}, \varepsilon, |\cdot|))=O\bigl(T(\log(1/\varepsilon)+nq)\bigr).
		\]
	\end{lemma}
        Note that the restriction of the exact \(q\)-bit representation can be relaxed by the \(q\)-bit approximation~\cite{jerbi2023quantum}.	We have the following corollary concerning the prediction error of such QML models:
	\begin{corollary}\label{app_corollary_2}
		(The prediction error of data re-uploading QML models)
		For the data re-uploading QML model defined in~\Cref{lemma_covering_re_uploading}, the expected prediction error of the optimal QML model on the training dataset $\tilde\boldf_{\hat{\boldtheta}_S}$ derived from the hypothesis space \(\mathcal{\tilde{H}}\) scales as \[\mathbb{E}_{\cP^{\otimes N}}\left[\cE_P(\tilde\boldf_{\hat{\boldtheta}_S}) \right]=\tilde{O}\left(nq\frac{T}{N}\right).\] 
	\end{corollary}
	\begin{proof}
		To prove this theorem, we only need to repeat the proof procedure of~\Cref{thm1_upper_bound}. Note that we should substitute the hypothesis space from \(\cH\) in~\Cref{thm1_upper_bound} to \(\mathcal{\tilde{H}}\) in this context. 
	\end{proof}
	
	\subsection{Extensions to other loss functions}
	In this study, the loss function is defined as the $\ell^2$ loss function $\ell(\boldf_{{\boldtheta}}; \bfx, y)=(\boldf_{{\boldtheta}}(\bfx)-y)^2$ for the sample pair $(\bfx, y)\in\cX\times\cY$. It is worth noting that this $\ell^2$ loss function can be extended to other loss functions that satisfy the property of Lipschitz continuity and strong convexity, ensuring the validity of Lemma~\ref{lem:cover} and Lemma~\ref{lem:Bernstein}. An illustrative example is the logistic regression loss function \(\ell(\boldf_{{\boldtheta}}; \bfx, y)=\log(1+\exp(-y\boldf_{{\boldtheta}}(\bfx)))\), which finds applications in classification problems.
	
	To assess the loss function values \(\ell(\boldf_{{\boldtheta}}; \bfx, y)\) on a quantum device, we need to measure the QML output $\boldf_{{\boldtheta}}(\bfx)$ and proceed it with \(y\) on a classical computer. However, a drawback of this evaluation process arises from the probabilistic nature of quantum measurements. Specifically, obtaining the exact value of $\boldf_{{\boldtheta}}(\bfx)$ is challenging in practical implementations. Therefore, it becomes necessary to approximate the output of the QML model by averaging the measurement outcomes. The calculation of the loss value, involving the difference between a non-deterministic QML model output and a deterministic value \(y\), is influenced by measurement errors. Moreover, the square operation in the \(\ell^2\) loss function propagates these errors, making it challenging to obtain accurate loss values. 
	
	An alternative approach, as demonstrated by \citet{cerezo2021CostFunctionDependent} and \citet{pan2023DeepQuantumNeural}, is to directly use the output of the QML model as the loss value. A common formulation of the loss functions from these works is 
	\begin{eqnarray}\label{app_eqn_general_loss}
		\tilde{\ell}(\boldtheta; \bfx, y)=\Tr\left[\cO_{\bfx, y}U_{\boldtheta}\rho(\bfx)U_{\boldtheta}^{\dagger}\right].
	\end{eqnarray}
	We note that Eq.~(\ref{app_eqn_general_loss}) only provides a simplified version of the original loss function. In more complex cases, the QML model $U_{\boldtheta}$ can be applied to the subsystem of $\rho(\bfx)$. The choice of the measurement observable $\cO_{\bfx, y}$ determines the specific purpose of the loss function. In the following discussion, we provide the reformulation of the \(\ell^2\) loss into Eq.~(\ref{app_eqn_general_loss}), which can be efficiently implemented on the quantum hardware.
	
	\paragraph{Trace distance between quantum states.}
	\citet{cerezo2021CostFunctionDependent} consider the state preparation problem, where the goal is to find a QML model \(U_{\boldtheta}\) applied on an input quantum state (say \(\ket{0}\)) that prepares a target state \(\ket{\psi_0}\). The loss function is defined to be dependent on the squared trace distance between \(\ket{\psi_0}\) and \(U_{\boldtheta}^{\dagger}\ket{0}\), i.e., \(|\bra{\psi_0}U_{\boldtheta}^{\dagger}\ket{0}|^2\). This loss function can be reformulated into Eq.~(\ref{app_eqn_general_loss}) by setting \(\rho(x)=\ket{\psi_0}\bra{\psi_0}\) and \(\cO_{x, y}=\bI-\ket{0}\bra{0}\), in which \(\bI\) is the identity matrix.
	
	We claim that the \(\ell^2\) distance between \(\ket{\psi_0}\) and \(U_{\boldtheta}^{\dagger}\ket{0}\) can also be reformulated into Eq.~(\ref{app_eqn_general_loss}). Concretely, we have 
	\begin{eqnarray}
		\ell(U_{\boldtheta}; \ket{0}, \ket{\psi_0})=\|U^{\dagger}\ket{0}- \ket{\psi_0}\|_2^2=\left(\bra{\psi_0}-\bra{0}U\right)\left(\ket{\psi_0}-U^{\dagger}\ket{0}\right) = 2-2\text{Re}(\bra{\psi_0}U^{\dagger}\ket{0}). \nonumber
	\end{eqnarray}
	The real part of the inner product \(\text{Re}(\bra{\psi_0}U^{\dagger}\ket{0})\) can be efficiently approximated based on quantum phase estimation (see \cite{zhao2019BuildingQNN} for reference). Consequently, we can formally construct a parameterized quantum circuit as our QML model, with the expectation of the measurement output being the \(\ell^2\) distance between \(\ket{\psi_0}\) and \(U_{\boldtheta}^{\dagger}\ket{0}\). Note that although our study analyzes the case where the response variable \(y\in\mathbb{R}\), our results can be generalized to \(\mathbb{R}^m\) case easily through component-by-component analysis, with an additional cost at most linearly dependent on the dimension \(m\). 
	
	\paragraph{Fidelity and Frobenius norm between quantum channels.}
	Additionally, \citet{pan2023DeepQuantumNeural} consider the quantum channel learning problem, aiming to learn an unknown quantum channel \(\mathcal{C}\) by a deep quantum neural network \(U_{\boldtheta}\). The training set composes of input quantum state \(\rho\) and output quantum state \(\tau=\mathcal{C}(\rho)\). When the target channel is a quantum unitary and the input states are pure states, they propose two loss functions based on the fidelity and Frobenius norm \(\|\cdot\|_F\). The fidelity loss function is defined as \(\Tr[U_{\boldtheta}\rho U_{\boldtheta}^{\dagger}\tau]\), where \(\rho\) is a pure state \(\ket{x}\bra{x}\) and \(\tau\) can be represented as \(\ket{y}\bra{y}\). Consequently, the fidelity between quantum channels can be reduced to the trace distance between quantum states as discussed earlier. For the Frobenius norm \(\|U_{\boldtheta}\rho U_{\boldtheta}^{\dagger}-\tau\|_F^2\), it is the generalization of the \(\ell^2\) loss to matrix space. Therefore, existing approaches to evaluating the loss value or loss function gradient in their work can be utilized to implement the \(\ell^2\) loss in this context.

\section{Proof of Theorem~2}\label{appen_thm2}	
    \subsection{The Gaussian Denoise Problem}
    We define a family of supervised learning problems using a linear QML
    model $(V(\bfx),U(\boldtheta),\cO)$, where $\cO$ has a bounded operator norm. For every $\boldtheta\in\Theta$, define the so-called Gaussian denoise problem associated with $\boldtheta$ as follows: Let $\bfx\in\cX$ be sampled according to an unknown density $\mu$. The label of $\bfx$ is given by 
    \[
    y_{\boldtheta}=\boldf_{\boldtheta}(\bfx)+\varepsilon,
    \]
    where $\varepsilon$ is an independent Gaussian noise with mean $0$ and bounded variance $\sigma^2\in(0,\infty)$ which is independent with $\bfx$. The goal is to recover $\boldtheta$ from a finite training data set $S=\{(\bfx_i, y_{\boldtheta, i})\}_{i=1}^{N}$ of size $N$, where the data are sampled i.i.d. with respect to $\mu$ and the Gaussian noise.
    Note that in this setting, the sample density is determined by the density $\mu$ and the Gaussian density: For any $\boldtheta\in\mathbf{\Theta}$, the density of the sample $z_{\boldtheta}=(\bfx, y_{\boldtheta})$ is given by
    \begin{equation}\label{sample distribution}
    \cP_{\boldtheta}(z_{\boldtheta})=\frac{1}{\sqrt{2\pi\sigma^2}}\exp{\frac{-(y-f_{\boldtheta}(\bfx))^2}{2\sigma^2}}\mu(\bfx).
    \end{equation}

    \subsection{The Minimax Risk of the Gaussian Denoise Problem}
    Let $\cA$ be any empirical QML strategy that takes a finite training data set $S$ and outputs a hypothesis function $\boldf_{\cA(S)}$ which approximates the target function $\boldf_{\boldtheta^*}$. To lower bound the prediction error $\cE_P(\boldf_{\cA(S)})$ for any QML strategy $\cA$ on the training data set $S$, we consider the following minimax problem:
 \begin{equation}\label{eq: minimax risk}
 \inf_{\cA} \sup_{\boldtheta} \mathbb{E}_{S\sim\cP_{\boldtheta}^{\otimes N}}\left[R(\boldf_{\mathcal A(S)})-R(\boldf_{\boldtheta})\right].    
 \end{equation}
 The supremum in~\cref{eq: minimax risk} evaluates the largest expected prediction error of an empirical QML strategy $\cA$ aiming to solve a Gaussian denoise problem associated with $\boldtheta$. If one can lower bound the expectation value of the minimax problem in~\cref{eq: minimax risk}, there exists a certain Gaussian denoise problem and a training data set $S$ satisfying that the expected prediction error of any empirical QML strategy on $S$ cannot be too small as well. 
 
We first note that for the Gaussian denoise problem, the prediction error is exactly the squared $L^2(\mu)$ distance between the estimator and the target function:
	\begin{eqnarray}
		~ &~& R(\boldf_{\mathcal A(S)})-R(\boldf_{\boldtheta}) \nonumber \\
		&=&  \mathbb{E}_{z_{\boldtheta}\sim\cP_{\boldtheta}}\left[(\boldf_{\mathcal A(S)}(\bfx)-y_{\boldtheta})^2-(\boldf_{\boldtheta}(\bfx)-y_{\boldtheta})^2\right]	\nonumber \\
		~ &=& \int_{\mathcal{X}} \int_{-\infty}^{\infty} \frac{1}{\sqrt{2\pi\sigma^2}}e^{-\frac{\varepsilon^2}{2\sigma^2}}\mu(\bfx)  \left((\boldf_{\mathcal A(S)}(\bfx)-\boldf_{\boldtheta}(\bfx)-\varepsilon)^2-(\boldf_{\boldtheta}(\bfx)-\boldf_{\boldtheta}(\bfx)-\varepsilon)^2 \right) d\varepsilon d\bfx \nonumber \\ 
		~ &=&\int_{\mathcal{X}}\mu(\bfx)\left(\boldf_{\mathcal A(S)}(\bfx)-\boldf_{\boldtheta}(\bfx)\right)^2 d\bfx \nonumber \\
		~ &=& \|\boldf_{\mathcal A(S)}-\boldf_{\boldtheta}\|_{L^2(\mu)}^2, \nonumber
	\end{eqnarray}
	where the third equality holds since the noise is sampled from an independent Gaussian distribution with mean $0$. 
 
 By Markov's inequality (cf.~\Cref{lem_markov_ineq}), we have 
 \[
 \mathbb{E}_{S\sim\cP_{\boldtheta}^{\otimes N}}\left[\|\boldf_{\cA(S)}-\boldf_{\boldtheta} \|_{L^2(\mu)}^2\right]\geq a\cdot\mathbb{P}_{S\sim\cP_{\boldtheta}^{\otimes N}}\left( \|\boldf_{{\mathcal A(S)}}-\boldf_{\boldtheta}\|_{L^2(\mu)}^2>a \right). 
 \]
 Thus, it is sufficient to analyze $\inf_{\cA} \sup_{\boldtheta} \mathbb{P}_{S\sim\cP_{\boldtheta}^{\otimes N}}\left( \|\boldf_{\mathcal A(S)}-\boldf_{\boldtheta} \|_{L^2(\mu)}^2>a \right)$ and we shall prove the following:
 
	\begin{theorem}\label{thm_minimax_lower_bound}
        Let $\mu$ be a density over $\bfx$ satisfying that there exists a constant $C_1\in(0,1)$ such that $\mu(\bfx)\geq C_1~\forall \bfx\in\mathcal{X}$. 
	Let $\ubar{\varepsilon} = \sqrt{{CT}/{N}}$ for some positive constant $C$. Then we have
		\begin{eqnarray}\label{eqn_minimax_1}
			\inf_{\cA} \sup_{\boldtheta} \mathbb{P}_{S\sim\cP_{\boldtheta}^{\otimes N}}\{ \|\boldf_{\mathcal A(S)}-\boldf_{\boldtheta} \|_{L^2(\mu)} > \ubar{\varepsilon}/2 \} \geq 1/2, 
		\end{eqnarray}
		Consequently, we have
		\begin{eqnarray}\label{eqn_minimax_2}
			\inf_{\cA} \sup_{\boldtheta} \mathbb{E}_{S\sim\cP_{\boldtheta}^{\otimes N}}[R(\boldf_{{\mathcal A(S)}})-R(\boldf_{\boldtheta})] \geq \frac{C}{8}\cdot\frac{T}{N}.
		\end{eqnarray}	
	\end{theorem}
\begin{proof}
Let $\mathcal{M}=\mathcal{M}(\mathcal{H}, \ubar{\varepsilon}, \|\cdot\|_{L^2(\mu)})$ and $\mathcal{N}=\mathcal{N}(\mathcal{H}, {\varepsilon}, \|\cdot\|_{L^2(\mu)})$ be the maximal $\ubar{\varepsilon}$-packing net and minimal $\varepsilon$-covering net of $\mathcal{H}$, respectively. We shall determine $\ubar{\varepsilon}$ and $\varepsilon$ later. It is clear that
\[
\sup_{\boldtheta} \bP_{S\sim\cP_{\boldtheta}^{\otimes N}}\{ \|{\boldf_{\mathcal A(S)}}-{\boldf_{\boldtheta}}\|_{L^2(\mu)} \geq \ubar{\varepsilon}/2 \} \geq \max_{\boldf_{\boldtheta}\in\mathcal{M}} \bP_{S\sim\cP_{\boldtheta}^{\otimes N}}\{ \|{\boldf_{\mathcal A(S)}}-{\boldf_{\boldtheta}}\|_{L^2(\mu)}\geq \ubar{\varepsilon}/2 \}.
\]
For the estimator $\boldf_{\mathcal A(S)}\in\cH$ obtained by an empirical QML strategy $\cA$ and training data set $S$, we pick the closest hypothesis function $\boldf_{\boldtheta_{\cA(S)}}$ from $\mathcal{M}$, i.e.~ 
\[
\boldf_{\boldtheta_{\cA(S)}}=\mathop{\arg\min}\limits_{\boldf_{\boldtheta}\in\mathcal{M}} \|\boldf_{\mathcal A(S)}-\boldf_{\boldtheta}\|_{L^2(\mu)}.
 \] 

Note that for any $\boldf_{\boldtheta}\in\mathcal{M}$, the condition $\boldf_{\boldtheta}\neq\boldf_{\boldtheta_{\cA(S)}}$ implies $\|{\boldf_{\mathcal A(S)}}-{\boldf_{\boldtheta}}\|_{L^2(\mu)}\geq \ubar{\varepsilon}/2$. Since $\|{\boldf_{\mathcal A(S)}}-{\boldf_{\boldtheta}}\|_{L^2(\mu)}< \ubar{\varepsilon}/2$ implies $\boldf_{\boldtheta}=\boldf_{\boldtheta_{\cA(S)}}$ by triangular inequality. Consequently, we have that for any QML strategy $\cA$,
		\begin{eqnarray}\label{appendix-lower-1}
            \sup_{\boldtheta}  \bP_{S\sim\cP_{\boldtheta}^{\otimes N}}\{ \|\boldf_{\mathcal A(S)}-\boldf_{\boldtheta}\|_{L^2(\mu)} \geq \ubar{\varepsilon}/2 \} 
             &\geq& \max_{\boldf_{\boldtheta}\in\mathcal{M}} \bP_{S\sim\cP_{\boldtheta}^{\otimes N}}\{ \|{\boldf_{\mathcal A(S)}}-{\boldf_{\boldtheta}}\|_{L^2(\mu)}\geq \ubar{\varepsilon}/2 \} \nonumber \\
             ~ &\geq& \max_{\boldf_{\boldtheta}\in\mathcal{M}} \bP_{S\sim\cP_{\boldtheta}^{\otimes N}}\{ \boldf_{\boldtheta}\neq \boldf_{\boldtheta_{\cA(S)}}\}. 
            \end{eqnarray}
Since $\mathcal{M}$ is a finite set, we can assign $\boldf_{\boldtheta}$ with the uniform distribution over $\mathcal{M}$. Eq.~(\ref{appendix-lower-1}) can be lower bounded by 
\begin{eqnarray}
    \max_{\boldf_{\boldtheta}\in\mathcal{M}} \bP_{S\sim\cP_{\boldtheta}^{\otimes N}}\{ \boldf_{\boldtheta}\neq   \boldf_{\boldtheta_{\cA(S)}}\} &\geq& \frac{1}{|\mathcal{M}|}\sum_{\boldf_{\boldtheta}\in\mathcal{M}}\bP_{S\sim\cP_{\boldtheta}^{\otimes N}}\{ \boldf_{\boldtheta}\neq \boldf_{\boldtheta_{\cA(S)}}\} \nonumber := \bP_{\boldf_{\boldtheta}\sim u, S\sim\cP_{\boldtheta}^{\otimes N}}\{ \boldf_{\boldtheta}\neq \boldf_{\boldtheta_{\cA(S)}}\},
\end{eqnarray}
where $\bP_{\boldf_{\boldtheta}\sim u, S\sim\cP_{\boldtheta}^{\otimes N}}$ denotes the Bayes average probability with respect to the uniform distribution $u$ over $\mathcal{M}$ and the conditional density $\cP_{\boldtheta}^{\otimes N}$.

At this moment, note that $\boldf_{\boldtheta} \rightarrow S \rightarrow \boldf_{\boldtheta_{\cA(S)}}$ forms a Markov chain, since the sample data $S=\{z_{\boldtheta, i}\}_{i=1}^N$ depends only on the choice of $f_{\boldtheta}$ and the estimator $\boldf_{\boldtheta_{\cA(S)}}$ only depends on the sample data set $S$. Using Fano's inequality (cf.~\Cref{Fano_ineq}), we have:
\begin{eqnarray}\label{appendix-lower-2}			\bP_{\boldf_{\boldtheta}\sim u, S\sim\cP_{\boldtheta}^{\otimes N}}\{ \boldf_{\boldtheta}\neq \boldf_{\boldtheta_{\cA(S)}}\}\geq 1-\frac{I(\boldf_{\boldtheta};~S)+\log 2}{\log(|\mathcal{M}|)}.
\end{eqnarray}
For the mutual information, we have
\begin{eqnarray}\label{appendix-lower-3}
    I(\boldf_{\boldtheta};~S) &=&\sum_{\boldf_{\boldtheta}\in\mathcal{M}} \int_{(\mathcal{X}\times \cY)^{\otimes N}} \frac{1}{|\mathcal{M}|}\cP_{\boldtheta}(z)^N \log(\frac{\cP_{\boldtheta}(z)^N}{\sum_{\boldf_{\boldtheta'}\in\mathcal{M}}1/|\mathcal{M}| \cP_{\boldtheta'}(z)^{N} })(dz)^{N} \nonumber \\
    &=& \sum_{\boldf_{\boldtheta}\in\mathcal{M}} \int_{(\mathcal{X}\times \cY)^{\otimes N}} \frac{1}{|\mathcal{M}|} \cP_{\boldtheta}(z)^N \log(\frac{\cP_{\boldtheta}(z)^N}{\cP_{\mathcal{M}}(z)^N })(dz)^{N}
\end{eqnarray}
where $\cP_{\mathcal{M}}(z)^N=\sum_{\boldf_{\boldtheta'}\in\mathcal{M}} \frac{1}{|\mathcal{M}|}\cP_{\boldtheta'}(z)^{N}$ describes the marginal distribution of the sample data $S$.

We further upper bound the mutual information using the following strategy: Consider the uniform distribution over $\mathcal{N}$. In this way, we define another marginal distribution of $S$ by $\mathcal{P}_{\mathcal{N}}(z)^N=\frac{1}{|\mathcal{N}|}\sum_{\boldf_{\boldtheta}\in\mathcal{N}}\cP_{\boldtheta}(z)^N$. Note that
\begin{eqnarray}
    ~ &~&\sum_{\boldf_{\boldtheta}\in\mathcal{M}} \int_{(\mathcal{X}\times \cY)^{\otimes N}} \frac{1}{|\mathcal{M}|}\cP_{\boldtheta}(z)^N \left(\log(\frac{\cP_{\boldtheta}(z)^N}{\cP_{\mathcal{N}}(z)^N})-\log(\frac{\cP_{\boldtheta}(z)^N}{\cP_{\mathcal{M}}(z)^N }) \right)(dz)^{N} \nonumber \\
    ~ &=& \sum_{\boldf_{\boldtheta}\in\mathcal{M}} \int_{(\mathcal{X}\times \cY)^{\otimes N}} \frac{1}{|\mathcal{M}|} \cP_{\boldtheta}(z)^N \log(\frac{\cP_{\mathcal{M}}(z)^N}{\cP_{\mathcal{N}}(z)^N}) (dz)^{N} \nonumber \\
    ~ &=& \int_{(\mathcal{X}\times \cY)^{\otimes N}} \cP_{\mathcal{M}}(z)^N \log(\frac{\cP_{\mathcal{M}}(z)^N}{\cP_{\mathcal{N}}(z)^N}) (dz)^{N} \nonumber \\
    ~ &=& \KL(\cP_{\mathcal{M}}(z)^N||\cP_{\mathcal{N}}(z)^N) \geq 0, \nonumber
\end{eqnarray}
where the last inequality uses the fact that the K-L divergence is nonnegative. Thus we have 
\begin{eqnarray}
    I(\boldf_{\boldtheta};~S) &\leq& \sum_{\boldf_{\boldtheta}\in\mathcal{M}} \int_{(\mathcal{X}\times \cY)^{\otimes N}} \frac{1}{|\mathcal{M}|}\cP_{\boldtheta}(z)^N \log(\frac{\cP_{\boldtheta}(z)^N}{\cP_{\mathcal{N}}(z)^N}) (dz)^{N} \nonumber \\
    ~ &\leq& \max_{\boldf_{\boldtheta}\in\mathcal{M}} \int_{(\mathcal{X}\times \cY)^{\otimes N}} \cP_{\boldtheta}(z)^N \log(\frac{\cP_{\boldtheta}(z)^N}{\cP_{\mathcal{N}}(z)^N}) (dz)^{N} \nonumber \\
    ~ &=& \max_{\boldf_{\boldtheta}\in\mathcal{M}} \KL(\cP_{\boldtheta}(z)^N||\cP_{\mathcal{N}}(z)^N).
\end{eqnarray}
Let $\boldf_{\tilde \boldtheta}$ be a maximizer of the above maximization problem. From the $\varepsilon$-covering net $\mathcal{N}$, we pick $\boldf_{\tilde \boldtheta'}\in\mathcal{N}$ such that
$\|\boldf_{\tilde \boldtheta}-\boldf_{\tilde \boldtheta'}\|_{L^2(\mu)}\leq \varepsilon$. It follows that 
\begin{eqnarray}\label{eqn_minimax2}
    \max_{\boldf_{\boldtheta}\in\mathcal{M}} \KL(\cP_{\boldtheta}(z)^N||\cP_{\mathcal{N}}(z)^N) &= &\KL\left(\cP_{\tilde \boldtheta}(z)^{N} ||\cP_{\mathcal{N}}(z)^N \right) \nonumber \\ 
    ~&=& \int_{(\mathcal{X}\times \cY)^{\otimes N}} \cP_{\tilde \boldtheta}(z)^{N} 
    \log\left(\frac{\cP_{\tilde \boldtheta}(z)^{N}}{\sum_{\boldf_{\boldtheta}\in\mathcal{N}}1/|\mathcal{N}|\cP_{\boldtheta}(z)^{N}} \right)(dz)^{N}  \nonumber \\
    ~ &\leq& \int_{(\mathcal{X}\times \cY)^{\otimes N}} \cP_{\tilde \boldtheta}(z)^{N}\log\left(\frac{\cP_{\tilde \boldtheta}(z)^{N}}{1/|\mathcal{N}|\cP_{\tilde \boldtheta'}(z)^{N}} \right)(dz)^{N}  \nonumber \\
    ~ &=& \int_{(\mathcal{X}\times \cY)^{\otimes N}} \cP_{\tilde \boldtheta}(z)^{N}\left(\log|\mathcal{N}|+\log\left(\frac{\cP_{\tilde \boldtheta}(z)^{N}}{\cP_{\tilde \boldtheta'}(z)^{N}} \right)\right)(dz)^{N} \nonumber \\
    ~ &=& \log(|\mathcal{N}|)+\int_{(\mathcal{X}\times \cY)^{\otimes N}} \cP_{\tilde \boldtheta}(z)^{N}\log\left(\frac{\cP_{\tilde \boldtheta}(z)^{N}}{\cP_{\tilde \boldtheta'}(z)^{N}} \right)(dz)^{N} \nonumber \\ 
    ~ &=& \log(|\mathcal{N}|)+\KL(\cP_{\tilde \boldtheta}(z)^{N}||\cP_{\tilde \boldtheta'}(z)^{N}). \nonumber
\end{eqnarray}
Below we shall show that this quantity $\KL(\cP_{\tilde \boldtheta}(z)^{N}||\cP_{\tilde \boldtheta'}(z)^{N})$ is relevant to the $L^2(\mu)$ distance $\|\boldf_{\tilde\boldtheta}-\boldf_{\tilde\boldtheta'}\|_{L^2(\mu)}^2$. Specifically, it holds that 
\begin{eqnarray}
    \log\left(\frac{\cP_{\tilde \boldtheta}(z)^{N}}{\cP_{\tilde \boldtheta'}(z)^{N}} \right) = N\log \left( \frac{\frac{1}{\sqrt{2\pi\sigma^2}}e^{\frac{-(y-f_{\tilde \boldtheta}(\bfx))^2}{2\sigma^2}}\mu(\bfx)}{\frac{1}{\sqrt{2\pi\sigma^2}}e^{\frac{-(y-f_{\tilde \boldtheta'}(\bfx))^2}{2\sigma^2}}\mu(\bfx)} \right)  = \frac{N}{2\sigma^2}\left((y-f_{\tilde \boldtheta'}(\bfx))^2-(y-f_{\tilde \boldtheta}(\bfx))^2 \right). \nonumber
\end{eqnarray}
Without loss of generality, we set $\sigma^2=\frac{1}{2}$ for the rest calculation. It follows that
\begin{eqnarray}
    ~ &~& \KL(\cP_{\tilde \boldtheta}(z)^{N}||\cP_{\tilde \boldtheta'}(z)^{N}) \nonumber \\
    ~ &=& \int_{(\mathcal{X}\times \cY)^{\otimes N}} \cP_{\tilde \boldtheta}(z)^{N}\log\left(\frac{\cP_{\tilde \boldtheta}(z)^{N}}{\cP_{\tilde \boldtheta'}(z)^{N}} \right)(dz)^{N} \nonumber \\
     ~ &=& N\int_{(\mathcal{X}\times \cY)^{\otimes N}} \left(\frac{1}{\sqrt{\pi}}e^{-{(y-\boldf_{{\tilde \boldtheta}}(\bfx))^2}}\mu(\bfx)\right)^N \left((y-\boldf_{\tilde \boldtheta'}(\bfx))^2-(y-\boldf_{\tilde \boldtheta}(\bfx))^2\right)(dy)^N(d\bfx)^N  \nonumber \\
    ~ &=& N\int_{(\mathcal{X}\times \mathcal{E})^{\otimes N}} \left(\frac{1}{\sqrt{\pi}}e^{-{\varepsilon^2}}\mu(\bfx)\right)^N N(\boldf_{\tilde\boldtheta}(\bfx)-\boldf_{\tilde\boldtheta'}(\bfx))^2 (d\varepsilon)^N (d\bfx)^N  \nonumber \\
    ~ &=&N \int_{\mathcal{E}^{\otimes N}} \left(\frac{1}{\sqrt{\pi}}e^{-{\varepsilon^2}}\right)^N (d\varepsilon)^N \int_{\mathcal{X}^{\otimes N}}\mu(\bfx)^N (\boldf_{\tilde\boldtheta}(\bfx)-\boldf_{\tilde\boldtheta'}(\bfx))^2 (d\bfx)^N  \nonumber \\
    ~ &=& N \int_{\mathcal{X}} \mu(\bfx) (\boldf_{\tilde\boldtheta}(\bfx)-\boldf_{\tilde\boldtheta'}(\bfx))^2 d\bfx \nonumber \\
    ~ &=&  N\|\boldf_{\tilde\boldtheta}-\boldf_{\tilde\boldtheta'}\|_{L^2(\mu)}^2 \nonumber \\
    ~ &\leq& N\varepsilon^2, 
\end{eqnarray}
where the third equality holds since the noise is sampled from a Gaussian distribution with mean $0$. It follows that \begin{eqnarray}\label{eqn_minimax222}
    I(\boldf_{\boldtheta};~S) \leq \max_{\boldf_{\boldtheta}\in\mathcal{M}} \KL\left(\cP(z)^{N} ||\cQ(z^N) \right) \leq \log(|\mathcal{N}|)+N\varepsilon^2. 
\end{eqnarray}
Using this uniform upper bound, we have that
\begin{eqnarray}
    \inf_{\cA} \sup_{\boldtheta} \mathbb{P}_{S\sim\cP_{\boldtheta}^{\otimes N}}\{ \|\boldf_{\mathcal A(S)}-\boldf_{\boldtheta} \|_{L^2(\mu)} > \ubar{\varepsilon}/2 \} &\geq& \inf_{\cA} \max_{\boldf_{\boldtheta}\in\mathcal{M}} \bP_{S\sim\cP_{\boldtheta}^{\otimes N}}\{ \boldf_{\boldtheta} \neq \boldf_{\boldtheta_{\cA(S)}}\} \text{ (by Eq.~(\ref{appendix-lower-1}))} \nonumber \\
    ~ &\geq& 1-\frac{I(\boldf_{\boldtheta};~S)+\log 2}{\log(|\mathcal{M}|)} \text{ (by Eq.~(\ref{appendix-lower-2}))} \nonumber \\
    ~ &\geq& 1-\frac{\log(|\mathcal{N}|)+N\varepsilon^2+\log2}{\log(|\mathcal{M}|)} \text{ (by Eq.~(\ref{eqn_minimax222}))}. \nonumber
\end{eqnarray}

To guarantee Eq.~(\ref{eqn_minimax_1}), we only need to determine \(\ubar{\varepsilon}\) and $\varepsilon$ satisfying
\begin{eqnarray}\label{eqn_minimax4}
    \log (|\mathcal{M}(\mathcal{H}, \ubar{\varepsilon}, \|\cdot\|_{L^2(\mu)})|)\geq 2\log(|\mathcal{N}(\mathcal{H}, {\varepsilon}, \|\cdot\|_{L^2(\mu)})|)+2N\varepsilon^2+2\log 2. 
\end{eqnarray}
The quantity $\log(|\mathcal{N}(\mathcal{H}, {\varepsilon}, \|\cdot\|_{L^2(\mu)})|)$ decreases monotonically as $\varepsilon$ gets larger, while $N\varepsilon^2$ increases monotonically as $\varepsilon$ gets larger. Following the logic of \citet{yang1999information}, we would like to choose $\varepsilon$ to satisfy 
\begin{eqnarray}\label{eqn_minimax3}
    \log(|\mathcal{N}(\mathcal{H}, {\varepsilon}, \|\cdot\|_{L^2(\mu)})|)=N\varepsilon^2.
\end{eqnarray}
From~\Cref{lemma:packing-number-L2norm} we know $\log(|\mathcal{N}(\mathcal{H}, {\varepsilon}, \|\cdot\|_{L^2(\mu)})|)=\Theta(T\log(1/\varepsilon))$ for the linear QML model with $T$ trainable parameters. To solve Eq.~(\ref{eqn_minimax3}), we define a continuous function \(g(\varepsilon)=N\varepsilon^2-T\log(1/\varepsilon)\). Assuming $N\geq 4 T$, we have that 
\begin{eqnarray}
    ~ &~& g(\sqrt{\frac{T}{N}}) = T + T\log(\sqrt{\frac{T}{N}}) \leq 0, \nonumber \\
    ~ &~& g(\sqrt{\frac{T}{N}\log_2(\frac{N}{T})})=T\log(\frac{N}{T}) + \frac{T}{2}\log(\frac{T}{N}\log(\frac{N}{T})) \geq 0, \nonumber
\end{eqnarray}
which implies that if $\varepsilon$ satisfies $g(\varepsilon)=0$, we have \(\sqrt{C'\frac{T}{N}}\leq \varepsilon\leq\sqrt{\frac{C'T}{N}\log(\frac{N}{C'T})}\) for some constant $C'$ depending on the prefactor of covering entropy. Since the packing entropy and covering entropy are of the same order for QML models with the same trainable parameters, we can find a constant $C\in(0, 1)$ such that $\ubar{\varepsilon}=\sqrt{C\frac{T}{N}}$ satisfying Eq.~(\ref{eqn_minimax4}), which completes the proof. 

\end{proof}

	\section{Numerical Simulations}\label{appendix:numerical_simulations}
	In this section, we provide the simulation details omitted in the main text.
	\subsection{Univariate function fitting}
	We have defined the target function and domain set in the main text. Here, we will provide more details about the simulation settings.
	
	\noindent\textbf{Sample data.} The random number seed for each round of sampling is fixed. For example, when we set $N=6$, we will uniformly sample $6$ points from the domain set. In the next round of tasks, we set $N=8$. Note that the $8$ sample points must contain the chosen $6$ sample points from the last round. The purpose of this setting is to reduce the disturbance caused by sample randomness in the experiment.
	
	\noindent\textbf{QML model, input state and measurement.} The QML model is constructed by the data re-uploading quantum circuit. Concretely, the quantum circuit is composed of $20$ single qubit blocks. Every block is composed of $4$ quantum gates ordered by $\RZ$ gate, $\RY$ gate, $\RZ$ gate, and $\RY$ gate. The first three gates are equipped with trainable parameters, and the last gate is used for data encoding. The input state is $\ket{0}$ state, and the output is measured by the observable Z. All the trainable parameters are initialized guided by the uniform distribution on $[-\pi/3, \pi/3]$. The loss function is fixed as the mean squared loss defined in the main text.
	
	\noindent\textbf{Optimization.} We use the Adam optimizer~\cite{KingBa15} to optimize the trainable parameters in the QML model with a learning rate of $0.001$. We set the hyperparameter $\beta_1$ and $\beta_2$ as $0.9$ and $0.999$. Here $\beta_1$ is the exponential decay rate for the first-moment estimate of the gradient. This parameter controls how much the past gradient information is incorporated into the current estimate. $\beta_2$  is the exponential decay rate for the second-moment estimate of the gradient. This parameter controls how much the past squared gradient information is incorporated into the current estimate. We stop the optimization process if the empirical loss on the training data is decreased below $0.001$. 
	
	\subsection{Classifying quantum states}
	\noindent\textbf{Sample data.} The data-generating scheme is the same as that used in the univariate function-fitting experiment.
	
	\noindent\textbf{QCNN model, input state and measurement.} Most construction details of the QCNN model in this work are inspired by \citet{congQuantumConvolutionalNeural2019}. Concretely, we use three sets of convolutional layers (one $4$-qubit convolution layer and three $3$-qubit convolution layers) followed by a single pooling layer (pooling groups of $3$ qubits down to $1$ qubit) and finally a fully connected layer to build the QCNN model. The readers are referred to Ref.~\cite{congQuantumConvolutionalNeural2019} for the definition of the convolutional layers, pooling layers, and fully connected layers. The input state is $\ket{0}$ state, and the output is measured by the observable X. All the trainable parameters are initialized guided by the uniform distribution on $[-\pi/2, \pi/2]$. The loss function is fixed as the mean squared loss defined in the main text. Here, we note that the QCNN code is based on the code in this link (\url{https://github.com/Jaybsoni/Quantum-Convolutional-Neural-Networks}).
	
	\noindent\textbf{Optimization.} Here, we use the gradient descent method to optimize the trainable parameters in the QCNN model. The gradient is calculated by the parameter shift rule. The learning rate is set as $10^5$. In every learning step, if the loss increases, we expand the learning rate by $5\%$. Otherwise, we decrease the learning rate by $50\%$. We update the parameters in the QCNN model based on the information from the gradient and learning rate. We use two metrics to terminate training and output the optimized QCNN model as an estimator of the optimal QML model on the training dataset. The first is if the relative loss between iterations is smaller $10^{-7}$. The second is a hard cap on the total number of iterations, which is set as $5000$ in our simulation.

\end{document}